\documentclass[electronic]{vgtc}             

\graphicspath{{figures/}{pictures/}{images/}{./}} 

\usepackage{times}                     
\usepackage{tabu}                      
\usepackage{booktabs}                  
\usepackage{lipsum}                    
\usepackage{mwe}                       
\usepackage{amsmath,amsfonts}
\usepackage{algorithmic}
\usepackage{algorithm}
\usepackage{array}
\usepackage[caption=false,font=normalsize,labelfont=sf,textfont=sf]{subfig}
\usepackage{textcomp}
\usepackage{stfloats}
\usepackage{url}
\usepackage{verbatim}
\usepackage{cite}
\usepackage{bbding}
\usepackage{amssymb}
\usepackage{soul}
\usepackage{makecell}
\usepackage{colortbl}
\usepackage{multirow}
\usepackage{makecell}
\definecolor{NavyBlue}{rgb}{0.0, 0.0, 0.5}

\usepackage{mathptmx}                  
\onlineid{10561}
\vgtccategory{Research}
\vgtcinsertpkg

\title{Taking Language Embedded 3D Gaussian Splatting into the Wild}

\author{Yuze Wang\thanks{e-mail: wangyuze1998@buaa.edu.cn} %
\and Yue Qi\thanks{e-mail: qy@buaa.edu.cn}\thanks{Corresponding Author} %
}
\affiliation{\scriptsize State Key Laboratory of Virtual Reality Technology and Systems, \\ School of Computer Science and Engineering, Beihang University}

\teaser{
  \centering
  \includegraphics[width=\linewidth]{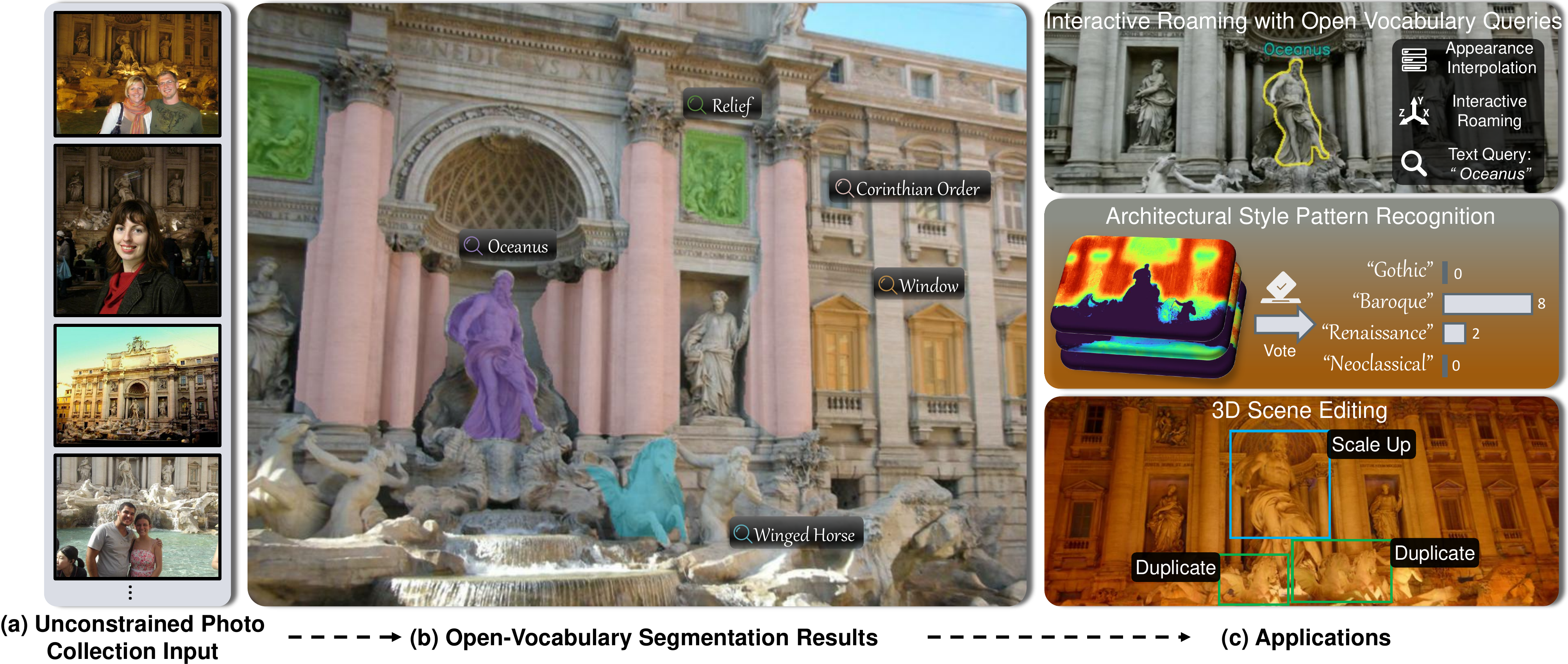}
  \caption{(a) Our method takes an unconstrained photo collection as input. (b) After optimization, our method achieve high-quality open-vocabulary segmentation. (c) The proposed method supports various applications, including interactive roaming with open-vocabulary queries, architectural style pattern recognition, and 3D scene editing.}
  \label{fig:teaser}
}

\abstract{
Recent advances in leveraging large-scale Internet photo collections for 3D reconstruction have enabled immersive virtual exploration of landmarks and historic sites worldwide.
However, little attention has been given to the immersive understanding of architectural styles and structural knowledge, which remains largely confined to browsing static text-image pairs.
Therefore, can we draw inspiration from 3D in-the-wild reconstruction techniques and use unconstrained photo collections to create an immersive approach for understanding the 3D structure of architectural components?
To this end, we extend language embedded 3D Gaussian splatting (3DGS) and propose a novel framework for open-vocabulary scene understanding from unconstrained photo collections.
Specifically, we first render multiple appearance images from the same viewpoint as the unconstrained image with the reconstructed radiance field, then extract multi-appearance CLIP features and two types of language feature uncertainty maps-transient and appearance uncertainty-derived from the multi-appearance features to guide the subsequent optimization process.
Next, we propose a transient uncertainty-aware autoencoder, a multi-appearance language field 3DGS representation, and a post-ensemble strategy to effectively compress, learn, and fuse language features from multiple appearances.
Finally, to quantitatively evaluate our method, we introduce PT-OVS, a new benchmark dataset for assessing open-vocabulary segmentation performance on unconstrained photo collections.
Experimental results show that our method outperforms existing methods, delivering accurate open-vocabulary segmentation and enabling applications such as interactive roaming with open-vocabulary queries, architectural style pattern recognition, and 3D scene editing.
Visit our project page at \href{https://yuzewang1998.github.io/takinglangsplatw/}{https://yuzewang1998.github.io/takinglangsplatw/}.
} 


\begin{document}

\newcommand{\energy}{E}
\newcommand{\mass}{m}
\newcommand{\lightspeed}{c}

\newcommand{\gs}{G_{k}(x)}

\newcommand{\gsCov}{\Sigma}
\newcommand{\gsTwoDCov}{\Sigma^{\prime}}
\newcommand{\gsPos}{\mu}
\newcommand{\gsScale}{S}
\newcommand{\gsRot}{R}
\newcommand{\gsOpa}{\alpha}
\newcommand{\gsFeat}{f}
\newcommand{\gsRadiance}{c}
\newcommand{\gsRadianceWild}{c}
\newcommand{\gsLangFeat}{lf}
\newcommand{\gsLangFeatApp}{lf}

\newcommand{\gsJacobian}{J}
\newcommand{\gsViewTransformation}{W}

\newcommand{\ray}{r}
\newcommand{\imgRenderRay}{\tilde{I}(r)}
\newcommand{\imgRender}{\tilde{I}}
\newcommand{\imgGT}{I}
\newcommand{\imgCLIPGT}{F}
\newcommand{\imgCLIPGTCompress}{H}
\newcommand{\imgCLIPRenderCompress}{\tilde{H}}
\newcommand{\MeanimgCLIPGT}{\bar{F}}
\newcommand{\imgCLIPRender}{\tilde{F}}
\newcommand{\imgGTCollection}{\mathbb{I}}

\newcommand{\LOne}{\mathcal{L}_{1}}
\newcommand{\LSSIM}{\mathcal{L}_{ssim}}
\newcommand{\LColor}{\mathcal{L}_c}
\newcommand{\LLang}{\mathcal{L}_{lang}}
\newcommand{\Lae}{\mathcal{L}_{ae}}

\newcommand{\appEmb}{l}
\newcommand{\transMask}{M}

\newcommand{\FwildAppEncoder}{f_{\alpha}}
\newcommand{\FwildTransientEncoder}{f_{\beta}}
\newcommand{\FTransferRadiance}{f_{\gamma}}
\newcommand{\FCLIP}{f_{clip}}
\newcommand{\FRenderer}{f_{render}}
\newcommand{\FEncoder}{f_{E}}
\newcommand{\FDecoder}{f_{D}}

\newcommand{\UncertainTrans}{U^{T}}
\newcommand{\UncertainApp}{U^{A}}

\newcommand{\gaussianRF}{\mathbb{G}_{RF}}

\newcommand{\textQuery}{Q_{text}}
\newcommand{\textNeg}{Q_{canon}}
\newcommand{\backgroundPredict}{{score}_{q}^{B}}
\newcommand{\score}{score}
\newcommand{\meanscore}{\bar{score}}

\newcommand{\chosenThre}{\tau}

\newcommand{\selConsQuality}{\epsilon_q}
\newcommand{\selConsDistance}{\epsilon_d}


\firstsection{Introduction}

\maketitle


Over the past few decades, researchers have used images collected from the Internet to reconstruct landmarks and historic sites, enabling their digital exploration and navigation \cite{snavely2011survey, snavely2006photo}.
However, when it comes to understanding architectural styles and structural knowledge, people still primarily rely on static text-image pairs, lacking immersive visual experiences.
\textit{Can we take inspiration from successful 3D in-the-wild reconstruction methods to achieve low-cost fine-grained 3D understanding of architectural components from Internet-sourced unconstrained photo collections?}
This could help users intuitively and immersively understand a building’s structure, style, and historical context.

Internet-sourced unconstrained photo collections often contain images captured under diverse conditions—across different years, lighting environments, camera devices, and viewpoints—and may include transient occluders such as pedestrians and vehicles moving through the scene.
While the understanding of architectural components through unconstrained photo collections is still in its infancy, recent advances in 3D in-the-wild reconstruction techniques offer valuable insights.

\begin{figure}[t]
\centering
\includegraphics[width=1.0\columnwidth]{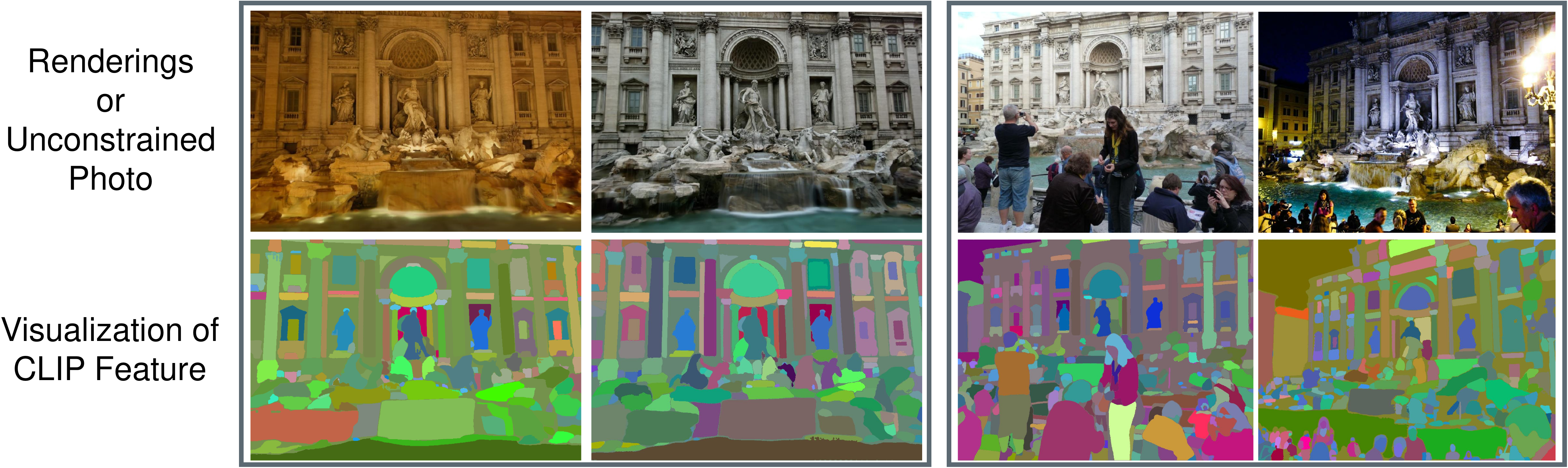}
\caption{
Left – Different appearances under the same viewpoint lead to significant variations in extracted CLIP features. Right – Transient occlusions present in the unconstrained image collection introduce unexpected CLIP features.
}
\label{fig:teaser_vis}
\end{figure}

Some studies have extended neural radiance fields (NeRFs) \cite{nerf_vanilla} and 3D Gaussian splatting (3DGS) \cite{3dgs} to unconstrained photo collection by predicting transient mask and latent appearance embedding for each unconstrained image.
NeRF-based methods \cite{nerf-w,ha-nerf,cr-nerf,k-planes} typically condition on appearance embeddings to generate image-specific radiance fields via a neural network, while most 3DGS-based approaches \cite{lookatthesky,we-gs,gs-w,xu2024wildgs,kulhanek2024wildgaussians,swag} achieve this by applying learnable affine transformations to the spherical harmonic (SH) coefficients for each 3D Gaussian.
These methods enable real-time rendering with smooth appearance and illumination changes of the landmarks and historic sites.
On the other hand, some 3DGS-based methods \cite{qin2024langsplat, feature3dgs, legaussians, goi} achieve open-vocabulary scene understanding using high-quality, well-captured photo collections. By learning 3D language embedded Gaussians with multi-view CLIP \cite{clip} features, they enable open-vocabulary segmentation.
Similar to in-the-wild 3D scene reconstruction, open-vocabulary scene understanding from unconstrained photo collections also faces challenges, such as appearance variations and transient occluders in each unconstrained image.
However, adapting these 3D open-vocabulary scene understanding or in-the-wild radiance field reconstruction methods to 3D understanding from unconstrained photo collections is not trivial:
First, due to the limited primary light sources in outdoor scenes—typically the sky and the sun—and the fact that appearance changes follow physical laws, modeling various appearances in a latent space is feasible. However, modeling latent language features (such as CLIP) in a similar manner is extremely challenging. Changes in language features can be highly noticeable and irregular due to variations in appearance, scale, occlusions, photo filters, and viewpoints.
The left part of \cref{fig:teaser_vis} visualizes how different appearances from the same viewpoint lead to varying language features, while the right part shows how transient occluders introduce unexpected language features.
Additionally, applying a black-and-white filter, for example, may implicitly introduce the semantics of an "old photograph," which is no value to the understanding of architectural components.
Secondly, prior in-the-wild 3D scene reconstruction methods assume that the radiance of a spatial point under varying appearances is additive, represented as an affine transformation of base radiance. However, this assumption does not hold for CLIP features due to their non-interpretability.
Furthermore, the semantics of architectural components should remain independent of appearance.
For instance, whether it is day or night, rain or shine, Oceanus remains at the center of the Trevi Fountain.

In this work, we take language embedded 3D Gaussian splatting from well-captured photo collections to unconstrained ones.
Specifically, we first render multiple appearance images from the same viewpoint as the unconstrained images, with the reconstructed in-the-wild radiance field, and extract multi-appearance CLIP features for post-ensemble.
Surprisingly, we find that fusing these pseudo-CLIP features with the original CLIP features can enhance open-vocabulary understanding performance.
Additionally, we extract two types of language feature uncertainty maps—appearance uncertainty map and transient uncertainty map—from the original and multi-appearance CLIP features.
These maps quantify semantic uncertainty caused by appearance variations and transient occluders, and are incorporated into language feature compression and language field reconstruction process.
To address storage constraints, we propose a transient uncertainty-awared autoencoder to compress the CLIP feature into a scene-specific lower-dimensional representation.
We further introduce a \underline{m}ulti-\underline{a}ppearance \underline{l}anguag\underline{e} 3DGS (MALE-GS) representation to learn the multi-appearance CLIP features. Then we propose a post-feature ensemble strategy to fuse the rendered multi-appearance CLIP features, enabling open-vocabulary queries.
Finally, while benchmarks and evaluation schemes exist for well-captured photo collections, none have been established for outdoor unconstrained scenes. Therefore, we introduce \underline{P}hoto \underline{T}ourism \underline{o}pen-\underline{v}ocabulary \underline{s}egmentation (PT-OVS) dataset, a new benchmark dataset derived from the public Photo Tourism \cite{snavely2006photo} dataset, providing dense annotations for seven scenes, including architectural components and their historical contexts, such as the "Rose Window" and "Last Judgment" in Notre-Dame de Paris.
Experimental results show that our method significantly outperforms state-of-the-art methods in open-vocabulary scene understanding, while enabling applications such as interactive roaming with open-vocabulary queries, architectural style pattern recognition, and 3D scene editing.

\section{Related Work}
\label{sec:2_related}

\subsection{Scene Representation and Radiance Fields}

Traditional methods have explored various 3D representations, such as volumes \cite{volume_1}, point clouds \cite{point_1,point_2}, meshes \cite{mesh_1}, and implicit functions \cite{if_1,if_2}, across a wide range of computer vision and graphics applications.
In recent years, neural radiance fields (NeRFs) \cite{nerf_vanilla} leveraged multi-layer perceptrons (MLPs) to model radiance fields, achieving high-fidelity novel view synthesis results.
A growing number of NeRF extensions emerged, targeting various goals such as enhancing visual effects \cite{mip-nerf,mip_nerf360,tri-miprf,wang2024scarf}, improving rendering speed \cite{mobile_nerf,merf}, and enabling scene editing \cite{rip-nerf,neumesh,nerf-editing,rise-editing}, among others.
3D Gaussian splatting (3DGS) \cite{3dgs} builded upon the NeRF concept by representing scenes as a set of 3D anisotropic Gaussians. This approach enabled real-time rendering and high-quality novel view synthesis while ensuring fast convergence.
Numerous studies further enhanced 3DGS, improving rendering quality \cite{gs_vis_mip-splatting,gs_vis_scaffold-gs,shen2025evolving}, reducing storage requirements \cite{gs_compact_1,gs_compact_2}, optimizing training processes \cite{gs_to_1,gs_to_2}, and exploring applications such as simultaneous localization and mapping (SLAM) \cite{gs_slam_gs-slam1,gs_slam_gs-slam2}, artifical intelligence generated content (AIGC) \cite{gs_aigc_dreamgs,qu2025drag,li2024director3d}, and scene understanding \cite{guo2025wildseg3d,legaussians,qin2024langsplat,goi,seg-wild}, among other applications.

\subsection{Novel View Synthesis for Unconstrained Photo Collections}
Although NeRF and 3DGS performed well with well-captured photo collections, they faced challenges when applied to unconstrained photo collections.
In real-world scenes, such as those in Internet photo collections \cite{snavely2006photo}, challenges arise not only from transient occluders, like moving pedestrians and vehicles, but also from varying illumination conditions.
NeRF-W \cite{nerf-w} was the first approach to apply NeRF for reconstructing scenes from unconstrained photo collections. It employed generative latent optimization (GLO) to learn appearance embeddings for each unconstrained image.
To render novel appearances from arbitrary images, Ha-NeRF \cite{ha-nerf} and CR-NeRF \cite{cr-nerf} employed convolutional neural networks (CNNs) to extract latent appearance embeddings for each input image.
To address the slow convergence of NeRF, K-Planes \cite{k-planes} introduced a volumetric NeRF representation that combined planar factorization with an MLP decoder. Recently, several studies explored replacing NeRF representations with 3DGS for this task. SWAG \cite{swag} proposed a learnable hash-grid latent appearance feature representation for appearance modeling. GS-W \cite{gs-w} introduced intrinsic and dynamic appearance features for each 3D Gaussian to handle variant appearances. WE-GS \cite{we-gs} introduced a lightweight spatial attention module that simultaneously predicted appearance embeddings and transient masks for each unconstrained image. SLS \cite{sabour2024spotlesssplats} and WildGaussians \cite{kulhanek2024wildgaussians} leveraged DINO \cite{dinov2} features extracted from each unconstrained image, which were then fed into a trainable neural network to predict transient occluders. Wild-GS \cite{xu2024wildgs} introduced hierarchical appearance decomposition and an explicit local appearance modeling strategy. Finally, Splatfacto-W \cite{xu2024splatfacto-w} implemented 3DGS for unconstrained photo collections within the NeRFstudio \cite{tancik2023nerfstudio} framework.

While these works have demonstrated promising results in in-the-wild scene radiance field reconstruction, few have explored the use of unconstrained photo collections for scene understanding.
\subsection{Open-Vocabulary 3D Scene Understanding}
Open-vocabulary scene understanding is a fundamental task in computer graphics and computer vision. Bridging 3D representations with natural language descriptions is crucial for enabling various applications, such as visual question answering, semantic segmentation, and object localization.
Leveraging vision-language models (VLMs) like CLIP \cite{clip}, methods such as CLIP2Scene \cite{clip2scene} and OpenScene \cite{openscene} developed point cloud-based language-embedded representations. With the advancement of NeRF, several work \cite{lerf,3dovs,ffd} incorporated dense CLIP features extracted from multi-view images into NeRF-based scene representations.
Very recently, several work aimed to bridge 3DGS representations with natural language descriptions to improve convergence efficiency, including LangSplat \cite{qin2024langsplat}, LEGaussians \cite{legaussians}, GOI \cite{goi}, GS Grouping \cite{gaussian_grouping}, and Feature 3DGS \cite{feature3dgs}. Following these work, CLIP-GS \cite{clipgs} and FastLGS \cite{fastlgs} further improved the rendering and convergence efficiency for language-embedded 3DGS by combining semantic feature grid or 3D coherent self-training.

However, the aforementioned methods mainly focused on indoor scenes with well-captured photo collections. In contrast, we aim to extend 3D open-vocabulary scene understanding to outdoor environments with unconstrained photo collections, such as Internet-sourced photo collections of landmarks and historic buildings.

\section{Preliminaries and Challenges}
\label{sec:pre}
\subsection{3DGS in the wild}

3DGS \cite{3dgs} represents scenes with millions of 3D anisotropic Gaussians. Each Gaussian $\gs$ is parameterized by its covariance matrix $\gsCov_{k}$, center position $\gsPos_{k}$, opacity $\gsOpa_{k}$, and features $\gsFeat_{k}$. In vanilla 3DGS, $\gsFeat_{k}$ is the spherical harmonic (SH) coefficients $\gsRadiance_{k}$ to model view-dependent appearance.
Given the Jacobian matrix $\gsJacobian$ of the affine projective transformation and the viewing transformation $\gsViewTransformation$, the 2D covariance matrix $\gsTwoDCov_{k}$ is computed as:
\begin{equation}
\gsTwoDCov_{k} = \gsJacobian \gsViewTransformation \gsCov_{k} \gsViewTransformation^{T} \gsJacobian^{T} .
\end{equation}
These 3D Gaussians are projected into 2D splats and the color of a pixel $\ray$, denoted as $\imgRenderRay$, is computed as follows:
\begin{equation}
\label{eq_vr}
\imgRenderRay = \sum_{i \in N} T_i \alpha_{i}^{\prime}f_{i}, \quad T_{i} = \prod_{j=1}^{i - 1}(1 - \alpha_{j}^{\prime}).
\end{equation}
The term $\gsOpa^{\prime}_k$ represents the final multiplied opacity of $\gsOpa_k$.
The attributes of each 3D Gaussian are optimized using the L1 loss $\LOne$ and the structural similarity index (SSIM) \cite{ssim} loss $\LSSIM$ between the rendered image $\imgRender$ and the ground truth (GT) image $\imgGT$.

To adapt 3DGS to unconstrained photo collections, some work \cite{we-gs,lookatthesky,kulhanek2024wildgaussians,gs-w} predict the appearance embedding $\appEmb_{i} \in \mathbb{R}^{d_a}$ and transient mask $\transMask_{i} \in \mathbb{R}^{W_t \times H_t}$ for each image $\imgGT_{i}$:
\begin{equation}
     \appEmb_{i}, \transMask_{i} = \FwildAppEncoder(\imgGT_{i}), \FwildTransientEncoder(\imgGT_{i}).
\end{equation}
Typically, $\FwildAppEncoder(.)$ and $\FwildTransientEncoder(.) $ are learnable CNNs. With injecting the appearance embedding $\appEmb_{i}$ into each 3D Gaussian, the appearance-specific radiance of each 3D Gaussian can be obtained with another learnable neural network $\FTransferRadiance$:
\begin{equation}
    \gsRadianceWild_{ik} =  \gsRadiance_{k} + \FTransferRadiance(\gsRadiance_{k},\gsPos_{k},\appEmb_{i}).
\end{equation}
By blending each 3D Gaussian with $\gsRadianceWild_{ik}$ using \cref{eq_vr}, an appearance-specific rendered image $\imgRender_{i}$ is obtained. The additional supervision is provided by the predicted transient mask $\transMask_{i}$:
\begin{equation}
    	\begin{aligned}
		\LColor &=  \LOne( (1-\transMask_{i}) \odot \imgRender_{i},(1-\transMask_{i}) \odot \imgGT_{i} ) \\
		&\quad+ \lambda \mathcal{L}_{SSIM}( (1-\transMask_{i}) \odot \imgRender_{i},(1-\transMask_{i}) \odot \imgGT_{i}  ),
	\end{aligned}
\end{equation}
where $\odot$ represents the Hadamard product and $\lambda$ is hyperparameter.
\subsection{Language Embedded 3DGS}
Some work \cite{goi,gaussian_grouping,feature3dgs,leg-gs} achieve open-vocabulary understanding with 3DGS. They first extract pixel-level CLIP features from each image:
\begin{equation}
    \imgCLIPGT_{i} = \FCLIP(\imgGT_{i}).
\end{equation}
By associating each 3D Gaussian $\gs$ with additional language features $\gsLangFeat_{k}$, rendering the feature map $\imgCLIPRender_{i}$ using the tile-based rasterizer with \cref{eq_vr}, and supervising the rendered language feature map with 2D language feature $\imgCLIPGT_{i}$, a language embedded 3DGS can be reconstructed:
\begin{equation}
    \LLang = \| \imgCLIPRender_{i} - \imgCLIPGT_{i} \|.
\end{equation}
However, these language embedded 3DGS methods struggle with unconstrained photo collections due to inconsistencies in multi-view CLIP features caused by varying appearances and transient occlusions. Moreover, the non-additive nature of CLIP features prevents the direct adaptation from in-the-wild radiance field reconstruction methods to in-the-wild language field reconstruction.

\begin{figure*}[!htb]
\centering
\includegraphics[width=2.0\columnwidth]{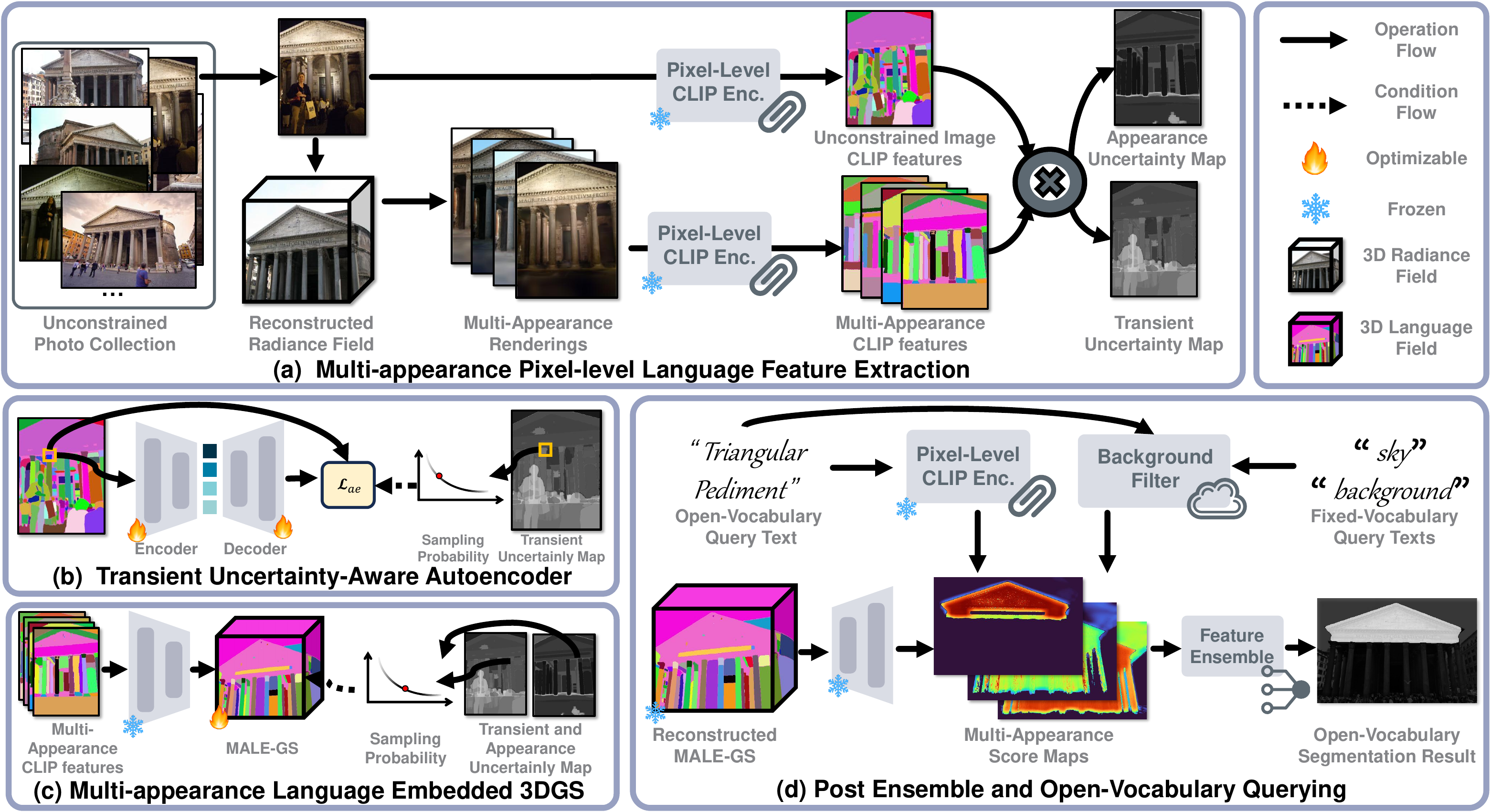}
\caption{Overview of our framework. (a) Given an image from an unconstrained photo collection, multi-appearance renderings are generated using the reconstructed radiance field. CLIP features are then extracted from both the unconstrained image and renderings using a pixel-level CLIP encoder. The language feature appearance and transient uncertainty map are derived from the extracted features. (b) A transient uncertainty-aware autoencoder is learned to compresses and restores the dense CLIP features. (c) With the MALE-GS representation, multi-appearance CLIP features are propagated into the 3D field, with uncertainty maps serving as optimization constraints. (d) Post-ensemble and background filter modules ensure high-quality open-vocabulary segmentation results.}
\label{fig:main}
\end{figure*}

\section{Method}
\label{sec:method}
Given a set of posed unconstrained images $\imgGTCollection = \{\imgGT_{1},\imgGT_{2},...\imgGT_{T}\}$ and a 3D Gaussian radiance field $\gaussianRF$ of the scene constructed by in-the-wild 3DGS method, such as WE-GS \cite{we-gs}. Our method expands $\gaussianRF$ with open-vocabulary semantics, enabling text-based queries.
\cref{fig:main} depicts the framework of our proposed method.
\subsection{Multi-Appearance Pixel-Level Language Feature Extraction}
\label{sec:method:feature_extraction}
Prior methods typically extract pixel-level CLIP features from images using either hierarchical SAM-based methods \cite{qin2024langsplat} or hierarchical cropped-image patch-based methods \cite{lerf, legaussians}.
Due to the scale variations in unconstrained images, we adopt a similar SAM-based approach, first applying SAM \cite{sam} to segment the image into distinct regions. Each segmented region is then cropped and processed through the CLIP image encoder to extract its corresponding CLIP features. Following LangSplat \cite{qin2024langsplat}, we extract features at three semantic levels: subpart, part, and whole.
To mitigate the inconsistency in illumination across unconstrained images, we generate multi-appearance CLIP features by first pre-rendering with multiple appearances and then encoding them with CLIP, as follows:
\begin{equation}
   \imgRender_{i}^{j} =\FRenderer(\imgGT_{i}, \FwildAppEncoder(\imgGT_{j}), \gaussianRF),
\end{equation}
and
\begin{equation}
   \imgCLIPGT_{i}^{j} =\FCLIP(\imgRender_{i}^{j}),
\end{equation}
where $\imgRender_{i}^{j}$ represents the rendered image at the camera pose of $\imgGT_{i}$, with its appearance conditioned on $\imgGT_{j}$.
In total, we render $N-1$ novel appearance images, along with a self-appearance rendering ($i = j$). We denote the set of multi-appearance pixel-level CLIP features for $\imgGT_{i}$ as: $\{\imgCLIPGT_{i}^{1}, \dots, \imgCLIPGT_{i}^{N-1}, \imgCLIPGT_{i}^{i}\}$.
We propose two constraints for selecting $N-1$ novel appearances for each scene. The first is the rendering quality constraint, which requires that the L1 distance between the rendered image and the source image of the selected appearance be lower than $\selConsQuality$:
\begin{equation}
    \|\FRenderer(\imgGT_{j}, \FwildAppEncoder(\imgGT_{j}), \gaussianRF) - \imgGT_{j} \| < \selConsQuality.
    \end{equation}
This ensures the quality of the rendered image, further maintaining the reliability of the CLIP features. Additionally, to avoid overly similar appearances, we introduce an appearance embedding distance constraint, which requires the Manhattan distance $Manhat(.,.)$ between the appearance embeddings of any two selected novel appearance images, $\imgGT_{j}$ and $\imgGT_{k}$, to exceed $\selConsDistance$:
\begin{equation}
    Manhat(\FwildAppEncoder(\imgGT_{j}) , \FwildAppEncoder(\imgGT_{k}) ) > \selConsDistance.
    \end{equation}
Additionally, we introduce two types of language feature uncertainty: appearance uncertainty and transient uncertainty. The appearance uncertainty map quantifies the variation in the CLIP features across multiple appearances:
\begin{equation}
\UncertainApp_i = \frac{1}{N}(\sum_{j=1}^{N-1}(\imgCLIPGT_{i}^{j} - \MeanimgCLIPGT_i)^2+(\imgCLIPGT_{i}^{i} - \MeanimgCLIPGT_i)^2),
\end{equation}
where $\MeanimgCLIPGT_i$ represents the mean of the multi-appearance CLIP features.
The transient uncertainty map is designed to measure the likelihood of transient occluders at language feature level:
\begin{equation}
\UncertainTrans_i = (\imgCLIPGT_i^i - \imgCLIPGT_i)^2,
\end{equation}
where $\imgCLIPGT_i^i$ represents the language feature extracted from the self-appearance rendering, and $\imgCLIPGT_i$ is the language feature extracted from the original unconstrained image.
Note that $\imgCLIPGT_i^i$ is only used for uncertainty calculation and will not be used in the later process.
\subsection{Transient Uncertainty-Aware Autoencoder}
\label{sec:method:autoencoder}
Due to the significant demands of storing high-dimensional language features for each 3D Gaussian, we pre-train an autoencoder \cite{ae} to compress the CLIP features into a lower-dimensional representation. These compressed features are then learned by the 3D language field, and decoded back to high-dimensional space during the evaluation stage.
However, the proposed multi-appearance CLIP features introduce $N-1$ times feature maps to compress.
Training the autoencoders on all the extracted CLIP features across different appearances significantly increases training time.
Moreover, we observe that fusing the features during the preprocessing (e.g., by averaging CLIP features across appearances) before training the autoencoder leads to a notable performance drop (as shown in \cref{tab:ablation_autoencoder}). We attribute this degradation to the expanded domain of definition of scene-specific CLIP features.

From our experiments, we identified a simple yet effective method: using only the CLIP features extracted from the original unconstrained images for autoencoder training. This strategy results in better segmentation performance. We believe this approach helps narrow the domain gap between the decompressed language feature maps of the rendered images and the original images. After compressing and decompressing the language features of the rendered image, they can be more closely aligned with the features extracted from the original image.

Besides, we introduce a transient uncertainty-aware optimization strategy when training of the autoencoder:
\begin{equation}
    \Lae = \Sigma_{t=1}^{T}\|\FDecoder(\FEncoder(\imgCLIPGT_t \odot (1-\UncertainTrans_t)))-\imgCLIPGT_t \odot (1-\UncertainTrans_t) \|,
\end{equation}
where $T$ is the number of unconstrained images, an encoder $\FEncoder$ maps the $D$-dimensional CLIP features $\imgCLIPGT_t$ to a $C$-dimensional representation. ($D$ is 512 in vanilla CLIP, and we set $C$ to 3 in our experiments.)
Additionally, to further accelerate autoencoder training, we set a condition where pixels with transient uncertainty values larger than $\tau_u$(set to 0.9 in our implementation) are assumed to be transient occluders and excluded from optimization.
\subsection{Multi-Appearance Language Embedded 3DGS}
\label{sec:method:male-gs}
After training the autoencoder, we transform all multi-appearance CLIP features $\{\imgCLIPGT_t^1, \dots,\imgCLIPGT_t^{N-1}, \imgCLIPGT_t\}$ into scene-specific compressed multi-appearance language features
$\{\imgCLIPGTCompress_t^1,\dots, \imgCLIPGTCompress_t^{N-1}, \imgCLIPGTCompress_t\}$.
To learn the multi-appearance language feature field, we propose the \underline{m}ulti-\underline{a}ppearance \underline{l}anguag\underline{e} 3DGS (MALE-GS) representation. Each 3D Gaussian is assigned $N$ language features $\{\gsLangFeatApp_1,..., \gsLangFeatApp_{N-1}, \gsLangFeatApp\}$ to learn $N-1$ novel appearances and a self-appearance language features.
In our experiments, we set $N = 4$. To maintain rendering efficiency, we also adopt a tile-based rasterizer with \cref{eq_vr} to render multi-appearance language features.

We optimize the 3D language field using the following transient-aware and appearance-aware language feature distance loss:
{\small
\begin{equation}
    	\begin{aligned}
		\LLang &= \sum_{t=1}^{T}(\|\imgCLIPGTCompress_t \odot (1-\UncertainApp_t) \odot (1-\UncertainTrans_t) - \imgCLIPRenderCompress_t \odot (1-\UncertainApp_t) \odot (1-\UncertainTrans_t) \|+ \\
		&\quad \sum_{n=1}^{N-1}\|\imgCLIPGTCompress_t^n \odot (1-\UncertainApp_t) \odot (1-\UncertainTrans_t) - \imgCLIPRenderCompress_t^n \odot (1-\UncertainApp_t) \odot (1-\UncertainTrans_t)\| ).
	\end{aligned}
\end{equation}
}
Here, the transient uncertainty map prevents transient occluders from contaminating the 3D Gaussian language field, while the appearance uncertainty map suppresses gradient updates for 3D Gaussians with significant language feature ambiguity across different appearances.
With other parameters of each MALE-GS fixed (e.g., position $\gsPos$, opacity $\gsOpa$), we only optimize the multi-appearance language features $\{\gsLangFeatApp_1,..., \gsLangFeatApp_{N-1}, \gsLangFeatApp\}$. For more details, please refer to the supplementary material.

\begin{table*}[]
    \caption{Quantitative experimental results on the proposed PT-OVS dataset, with the first, second, and third values highlighted in red, orange, and yellow, respectively. Our method demonstrates superior overall performance compared to state-of-the-art approaches.}
    \label{tab:quantitative_compare}
    \centering
    \LARGE
    \resizebox{2.0\columnwidth}{!}{%
\begin{tabular}{lccccccccccccccccccccc}
    \cmidrule[\heavyrulewidth]{2-22}
                  & \multicolumn{3}{c}{Brandenburg Gate}                                                                        & \multicolumn{3}{c}{Trevi Fountain}                                                                        & \multicolumn{3}{c}{Todaiji Temple}                                                                        & \multicolumn{3}{c}{Pantheon}                                                                        & \multicolumn{3}{c}{Taj Mahal}                                                                        & \multicolumn{3}{c}{Buckingham Palace}                                                                        & \multicolumn{3}{c}{Notre-Dame de Paris}                                                                        \\
    \cmidrule(rl){2-4} \cmidrule(rl){5-7} \cmidrule(rl){8-10} \cmidrule(rl){11-13} \cmidrule(rl){14-16} \cmidrule(rl){17-19} \cmidrule(rl){20-22}
                  & mIoU$\uparrow$                         & mPA$\uparrow$                           & mP$\uparrow$                            & mIoU$\uparrow$                          & mPA$\uparrow$                           & mP$\uparrow$                            & mIoU$\uparrow$                          & mPA$\uparrow$                           & mP$\uparrow$                            & mIoU$\uparrow$                          & mPA$\uparrow$                           & mP$\uparrow$                            & mIoU$\uparrow$                          & mPA$\uparrow$                           & mP$\uparrow$                            & mIoU$\uparrow$                          & mPA$\uparrow$                           & mP$\uparrow$                            & mIoU$\uparrow$                          & mPA$\uparrow$                           & mP$\uparrow$                            \\
    \midrule
LEGaussians \cite{legaussians}       & 0.158                         & 0.713                         & 0.161                         & OOM                            & OOM                           & OOM                            & 0.043                         & 0.402                         & 0.043                         & \cellcolor[HTML]{FFFCCC}0.356 & 0.644                         & \cellcolor[HTML]{FFFCCC}0.356 & \cellcolor[HTML]{FFFCCC}0.112 & \cellcolor[HTML]{FFFCCC}0.839 & \cellcolor[HTML]{FFFCCC}0.114 & \cellcolor[HTML]{FFFCCC}0.166 & \cellcolor[HTML]{FFFCCC}0.476 & \cellcolor[HTML]{FFFCCC}0.169 & OOM                            & OOM                            & OOM                           \\
Feature 3DGS \cite{feature3dgs}     & 0.028                         & 0.748                         & 0.029                         & OOM                            & OOM                            & OOM                            & 0.042                         & 0.648                         & 0.046                         & 0.078                         & \cellcolor[HTML]{FFFCCC}0.704 & 0.105                         & 0.069                         & 0.442                         & 0.073                         & 0.071                         & 0.386                         & 0.071                         & OOM                            & OOM                            & OOM                           \\
GS Grouping \cite{gaussian_grouping} & \cellcolor[HTML]{FCCC9C}0.456 & \cellcolor[HTML]{FCCC9C}0.961 & \cellcolor[HTML]{FCCC9C}0.539 & \cellcolor[HTML]{FFFCCC}0.010 & \cellcolor[HTML]{FFFCCC}0.583 & \cellcolor[HTML]{FFFCCC}0.010 & \cellcolor[HTML]{FFFCCC}0.052 & \cellcolor[HTML]{FCCC9C}0.906 & \cellcolor[HTML]{FFFCCC}0.053 & 0.009                         & 0.611                         & 0.009                         & 0.107                         & 0.818                         & 0.107                         & 0.105                         & 0.107                         & 0.105                         & \cellcolor[HTML]{FFFCCC}0.116 & \cellcolor[HTML]{FFFCCC}0.624 & \cellcolor[HTML]{FFFCCC}0.116 \\
LangSplat \cite{qin2024langsplat}        & \cellcolor[HTML]{FFFCCC}0.275 & \cellcolor[HTML]{FFFCCC}0.787 & \cellcolor[HTML]{FFFCCC}0.331 & \cellcolor[HTML]{FCCC9C}0.539 & \cellcolor[HTML]{FCCC9C}0.944 & \cellcolor[HTML]{FCCC9C}0.649 & \cellcolor[HTML]{FCCC9C}0.203 & \cellcolor[HTML]{FFFCCC}0.704 & \cellcolor[HTML]{FCCC9C}0.217 & \cellcolor[HTML]{FCCC9C}0.801 & \cellcolor[HTML]{FCCC9C}0.971 & \cellcolor[HTML]{FCCC9C}0.898 & \cellcolor[HTML]{FCCC9C}0.445 & \cellcolor[HTML]{FCCC9C}0.911 & \cellcolor[HTML]{FCCC9C}0.491 & \cellcolor[HTML]{FCCC9C}0.297 & \cellcolor[HTML]{FCCC9C}0.798 & \cellcolor[HTML]{FCCC9C}0.459 & \cellcolor[HTML]{FCCC9C}0.449 & \cellcolor[HTML]{FCCC9C}0.813 & \cellcolor[HTML]{FCCC9C}0.481 \\
    \midrule
Ours              & \cellcolor[HTML]{FFB2B2}0.619 & \cellcolor[HTML]{FFB2B2}0.985 & \cellcolor[HTML]{FFB2B2}0.846 & \cellcolor[HTML]{FFB2B2}0.593 & \cellcolor[HTML]{FFB2B2}0.969 & \cellcolor[HTML]{FFB2B2}0.757 & \cellcolor[HTML]{FFB2B2}0.321 & \cellcolor[HTML]{FFB2B2}0.940 & \cellcolor[HTML]{FFB2B2}0.381 & \cellcolor[HTML]{FFB2B2}0.928 & \cellcolor[HTML]{FFB2B2}0.990 & \cellcolor[HTML]{FFB2B2}0.977 & \cellcolor[HTML]{FFB2B2}0.613 & \cellcolor[HTML]{FFB2B2}0.967 & \cellcolor[HTML]{FFB2B2}0.693 & \cellcolor[HTML]{FFB2B2}0.540 & \cellcolor[HTML]{FFB2B2}0.925 & \cellcolor[HTML]{FFB2B2}0.668 & \cellcolor[HTML]{FFB2B2}0.736 & \cellcolor[HTML]{FFB2B2}0.926 & \cellcolor[HTML]{FFB2B2}0.892 \\
        \bottomrule
\end{tabular}
 }
\end{table*}

\subsection{Post Ensemble and Open-Vocabulary Querying}
After optimizing MALE-GS, open-vocabulary 3D queries can be performed with post-feature ensemble.
Given a specific viewpoint, the compressed multi-appearance language features are rendered. Then, using the trained decoder, we decode these compressed language features back into high-dimensional multi-appearance CLIP features $\{\imgCLIPRender^1,...,\imgCLIPRender^{N-1},\imgCLIPRender\}$. For readability, we denote $\imgCLIPRender$ as $\imgCLIPRender^{N}$.
Given a text query $\textQuery^q$, we first compute $N$ relevancy score map for each feature map with the following equation:
\begin{equation}
    \score_{q}^i = min_j\frac{exp(\textQuery^q \cdot \imgCLIPRender^i)}{exp(\textQuery^q \cdot \imgCLIPRender^i) + exp(\textNeg^{j} \cdot \imgCLIPRender^i)}.
\end{equation}
These score maps represent how much closer the rendered embedding is towards the query embedding compared to the canonical embeddings. All renderings use the same canonical texts $\{\textNeg^{j}|j=1,2,...,J\}$, such as "things" and "stuff". We find with such strategy, the background, especially the sky, will get a high score. So, we propose a backgroud filter to additional exclude the background, with the query texts "sky" and "background", and set the canonical texts as the actually query text $\textQuery^q$. We use $1$ minus its output as an additional score for subsequent fusion, referring to this as $\backgroundPredict$.

We find that the maximum value in the score map is positively correlated with the segmentation accuracy. Therefore, we use the maximum value of each score map as the weight to perform a weighted fusion of the multi-appearance score maps:
\begin{equation}
    \score_{q} = \sum_{i=1}^N \frac{\max(\score_{q}^i\cdot\backgroundPredict)}{\sum_{j=1}^N\max(\score_{q}^j\cdot\backgroundPredict)}\cdot\score_{q}^i\cdot\backgroundPredict.
\end{equation}
For open-vocabulary segmentation, we filter out pixels with fused relevancy scores lower than a predefined threshold $\chosenThre$ ($\chosenThre$ is 0.4 in our experiments).

\subsection{The PT-OVS Benchmark}
To evaluate our method, we require unconstrained photo collections paired with ground-truth open-vocabulary segmentation maps. However, to the best of our knowledge, no such dataset currently exists. Therefore, we introduce a new benchmark, PT-OVS, assembled from the Photo Tourism \cite{snavely2006photo} dataset. We selected seven scenes covering landmarks from various countries, architectural styles, and historical contexts.
NeRF-W \cite{nerf-w} introduced an unconstrained photo dataset for the Brandenburg Gate and Trevi Fountain to support in-the-wild radiance field reconstruction. They filtered out low-quality and heavily occluded images using NIMA \cite{talebi2018nima} and DeepLab V3 \cite{chen2017rethinking}. Following a similar approach, we curated a dataset for five additional scenes: Buckingham Palace, Notre-Dame de Paris, Pantheon, Taj Mahal, and Todaiji Temple.
We associate these landmarks with open-vocabulary descriptions, such as "Iron Cross" and "Bronze Sculpture" for Brandenburg Gate, and "Rose Window" and "Last Judgment" for Notre-Dame de Paris. We labeled 10–20 ground-truth segmentation maps for each scene based on selected high-quality unoccluded photos. More details about the proposed benchmark can be found in the supplementary material.

\begin{figure*}[!htb]
    \centering
    \includegraphics[width=1.92\columnwidth]{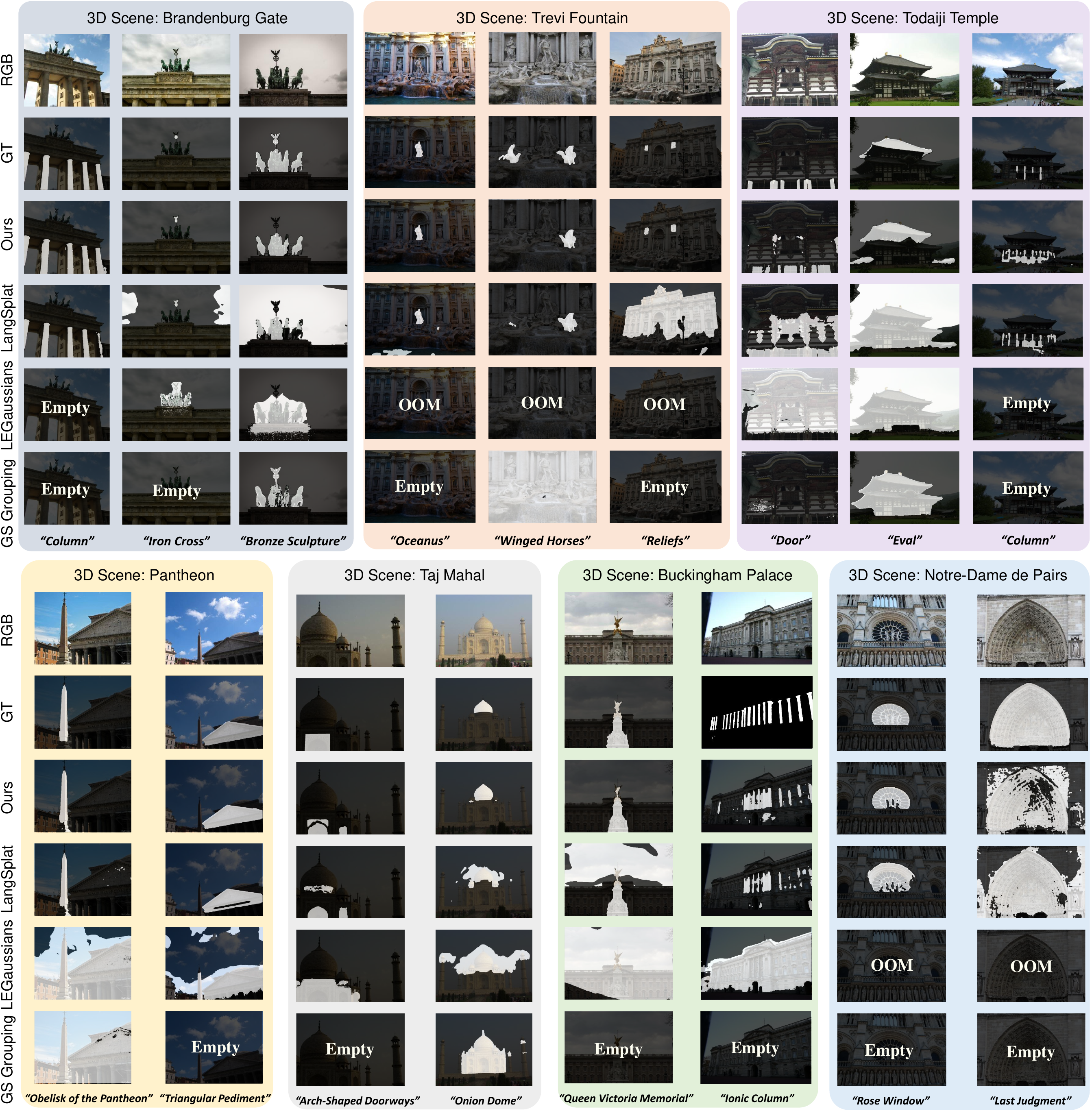}
    \caption{Qualitative experimental results on the proposed PT-OVS dataset, comparing our method with state-of-the-art approaches.}
    \label{fig:compare}
    \end{figure*}
    
    \section{Experiments}
    \label{sec:4_experiments}
    
    \subsection{Implementation Details}
    We use the SAM ViT-H model \cite{vit} for segmentation and the OpenCLIP ViT-B/16 model \cite{openclip} for CLIP feature extraction from each unconstrained image.
    We implement our method in Python using the PyTorch framework \cite{pytorch}, integrating custom CUDA acceleration kernels based on the differentiable Gaussian rasterization proposed by 3DGS \cite{3dgs}. We train the uncertainty-aware autoencoder for 100 epochs with a learning rate of 0.0001 and train the MALE-GS for 30,000 iterations with a learning rate 0.0025.
    All experiments are run on a single NVIDIA RTX-4090 GPU. For additional implementation details, please refer to the supplementary material.

    \subsection{Baselines and Metrics}
    We compare our approach to state-of-the-art 3DGS-based open-vocabulary scene understanding methods, including LangSplat \cite{qin2024langsplat}, Feature 3DGS \cite{feature3dgs}, GS Grouping \cite{gaussian_grouping}, and LEGaussians \cite{legaussians}. These methods focus on reconstructing language embedded 3DGS for static scenes from well-captured photo collections. We make minimal adjustments to adapt them to unconstrained images with varying resolutions.
    We quantitatively evaluate segmentation quality using mean intersection over union (mIoU), mean pixel accuracy (mPA), and mean precision (mP), comparing the predicted segmentation masks with the ground truth.
    \subsection{Comparisons}
    \subsubsection{Quantitative Comparison}
    
    The quantitative results are shown in \cref{tab:quantitative_compare}. Note that since all these methods use CLIP to extract language features from multi-view images, their level of "understanding of architectural knowledge" remains consistent. The differences in segmentation accuracy arise from their ability to robustly reconstruct the language embedded 3DGS.
    Specifically, compared to the second-best method, LangSplat, our approach achieves an average improvement of 19.1\% in mIoU across seven scenes. LEGaussians and Feature 3DGS encountered out-of-memory (OOM) issues in the Trevi Fountain and Notre-Dame de Paris scenes, preventing them from reconstructing the language embedded 3DGS. These two scenes each contain approximately 1,700 and 3,000 unconstrained images, respectively.
    Due to the varying scales of these images, the reconstructed scenes are rich in detail and contain a larger number of 3D Gaussians. In Feature 3DGS, each 3D Gaussian explicitly stores an uncompressed 512-dimensional CLIP features, making it impossible to load all 3D Gaussians into memory for optimization in these scenes. For LEGaussians, the dense quantization of language features for a large number of unconstrained images, along with additional learnable attributes per 3D Gaussian, also leads to OOM.
    
 As shown in \cref{tab:compare_eff_revised}, we compare the efficiency of our proposed method with state-of-the-art approaches. The training time encompasses the total duration of radiance field training, autoencoder training, and language field training. In terms of storage, our method introduces only an additional 40.6MB overhead compared to LangSplat. Furthermore, our method achieves state-of-the-art performance in both inference speed and accuracy. Overall, our approach achieves a better balance between efficiency and accuracy.
    \begin{table}[htb]

\caption{Comparison about the efficiency on the proposed PT-OVS dataset
\cite{pt}. Metrices are averaged over 5 scenes (except for Notre-Dame de Paris
and Trevi Fountain).
"TT", "IT", and "ST" are "training time ", "inference time", and "storage",
respectively.}
\label{tab:compare_eff_revised}
\resizebox{1.0\columnwidth}{!}{
\begin{tabular}{lccrr}
\cmidrule[\heavyrulewidth]{2-5}
                                       &TT (h)$\downarrow$ & IT (s) $\downarrow$ &ST (MB) $\downarrow$  & mIoU (\%) $\uparrow$ \\ \hline
LEGaussians \cite{legaussians}         & 3.15              & 0.18                & 362.8                & 16.7 \\
Feature 3DGS \cite{feature3dgs}        & 3.92              & 2.10                & 709.7                & 5.8 \\
GS Grouping \cite{gaussian_grouping}   & \textbf{0.81}     & 2.42                & 134.6                & 14.6 \\
LangSplat \cite{qin2024langsplat}      & 2.23              & \textbf{0.17}       & \textbf{74.4}        & 40.4 \\
WE-GS (Ours)                           & 3.08              & \textbf{0.17}       & 115.0                & \textbf{60.4} \\ \hline
\end{tabular}}
\end{table}

    \subsubsection{Qualitative Comparison}
    In \cref{fig:compare}, we visualize the segmentation results of both the baseline methods and our approach. Taking Brandenburg Gate as an example, our method accurately identifies architectural components (e.g., "Column"), components materials (e.g., "Bronze Sculpture"), and even the background knowledge associated with the architecture (e.g., "Iron Cross").
    Other baseline methods directly extract pixel-level CLIP features from images captured under varying appearances and do not account for the exclusion of occluders. As a result, the extracted multi-view CLIP features lack 3D consistency, making accurate 3D language field reconstruction difficult.
    
\begin{figure}[!htb]
        \centering
        \includegraphics[width=1.0\columnwidth]{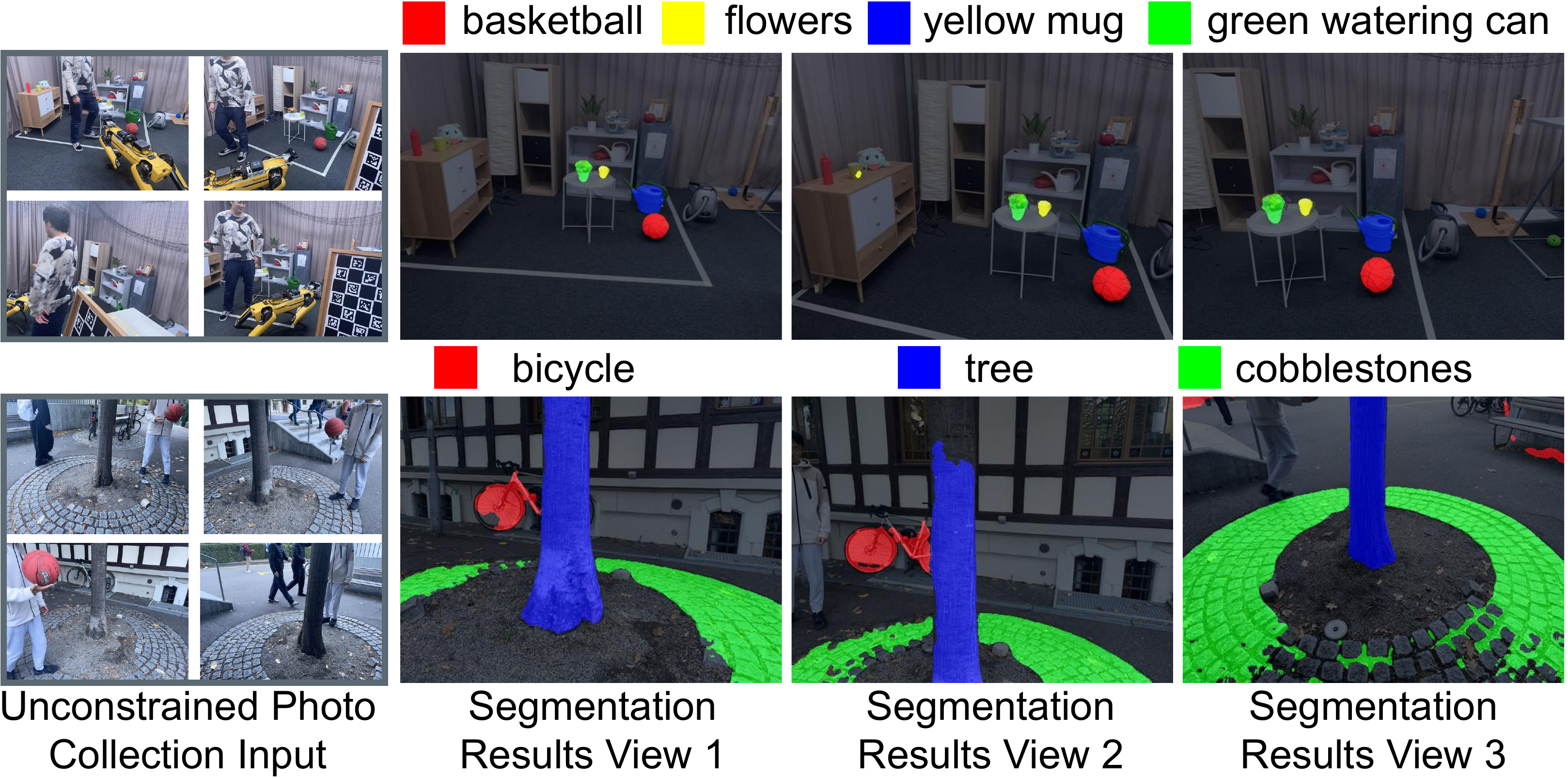}
        \caption{Qualitative experimental results on the NeRF-on-the-go dataset \cite{ren2024nerf-on-the-go}, demonstrating the generalization ability of our method.}
        \label{fig:otg}
        \end{figure}

    To further demonstrate the generalization ability of our method, we conduct experiments on two scenes from the NeRF-on-the-go dataset \cite{ren2024nerf-on-the-go}. As illustrated in \cref{fig:otg}, our method successfully reconstructs language-embedded 3DGS in both indoor and outdoor natural scenes with unconstrained photo collections.

    \subsection{Ablation Study}
    We validate the design choices of the proposed method on all scenes of the PT-OVS dataset. \cref{ablation_study}, \cref{tab:ablation_study_app}, and \cref{tab:ablation_autoencoder} present the quantitative results, while \cref{fig:transvis} and \cref{fig:fig_ablation_rfrecon} present the qualitative results.

\begin{table}[htbp]
\caption{Ablation studies of our method. Metrics are averaged over 7 scenes in the proposed PT-OVS dataset.}
\resizebox{1.0\columnwidth}{!}{
\begin{tabular}{clccc}
\hline
  \multicolumn{2}{c}{}       & mIoU$\uparrow$           & mPA$\uparrow$   & mP$\uparrow$               \\ \hline
\multirow{5}{*}{Preprocess} &(1) w/o Multi-Appearance Enh.  & 0.481     & 0.877          & 0.570                \\
                             & (2) w/o Post Ensemble                  & 0.502     & 0.890          & 0.613            \\
                             & (3) w/o Uncertainly-aware $\Lae$
                             & 0.529     & 0.906          & 0.630            \\
                             & (4) $\tau_u$ = 0.85 & 0.613     & 0.953          & 0.742            \\
                             & (5) $\tau_u$ = 0.95 & 0.620     & 0.956          & \textbf{0.745}            \\\hline

                             \multirow{3}{*}{\makecell{Language Field \\ Learning}}&(6) w/o TUM \& AUM        & 0.612     & 0.955          & 0.733            \\
                             &(7) w/o TUM      & 0.620       & \textbf{0.957}         & 0.743               \\
                             &(8) w/o AUM      & 0.612       & 0.954          & 0.735                  \\\hline
\multirow{5}{*}{Evaluation}&(9) w/o Bkg. filter      & 0.608      & 0.952         & 0.731                    \\
                      &(10) ImgLvlMax. Ens.          & 0.590         & 0.952        & 0.698 \\ 
                      &(11) PixMax. Ens.          & 0.564         & 0.949       & 0.650 \\   %
                      &(12)  PixAvg. Ens.         & 0.621         & 0.956        & 0.651 \\   %
                      &(13) PixWeightedAvg. Ens.          & 0.563         & 0.949        & 0.648\\\hline
{Ours}&(14) Completed Model    & \textbf{0.621}      & \textbf{0.957}         & \textbf{0.745}                   \\ \hline
\end{tabular}
}
\label{ablation_study}
\end{table}

    \subsubsection{The Influence of Multi-Appearance Enhancement}

\begin{figure}[!htb]
    \centering
    \includegraphics[width=1.0\columnwidth]{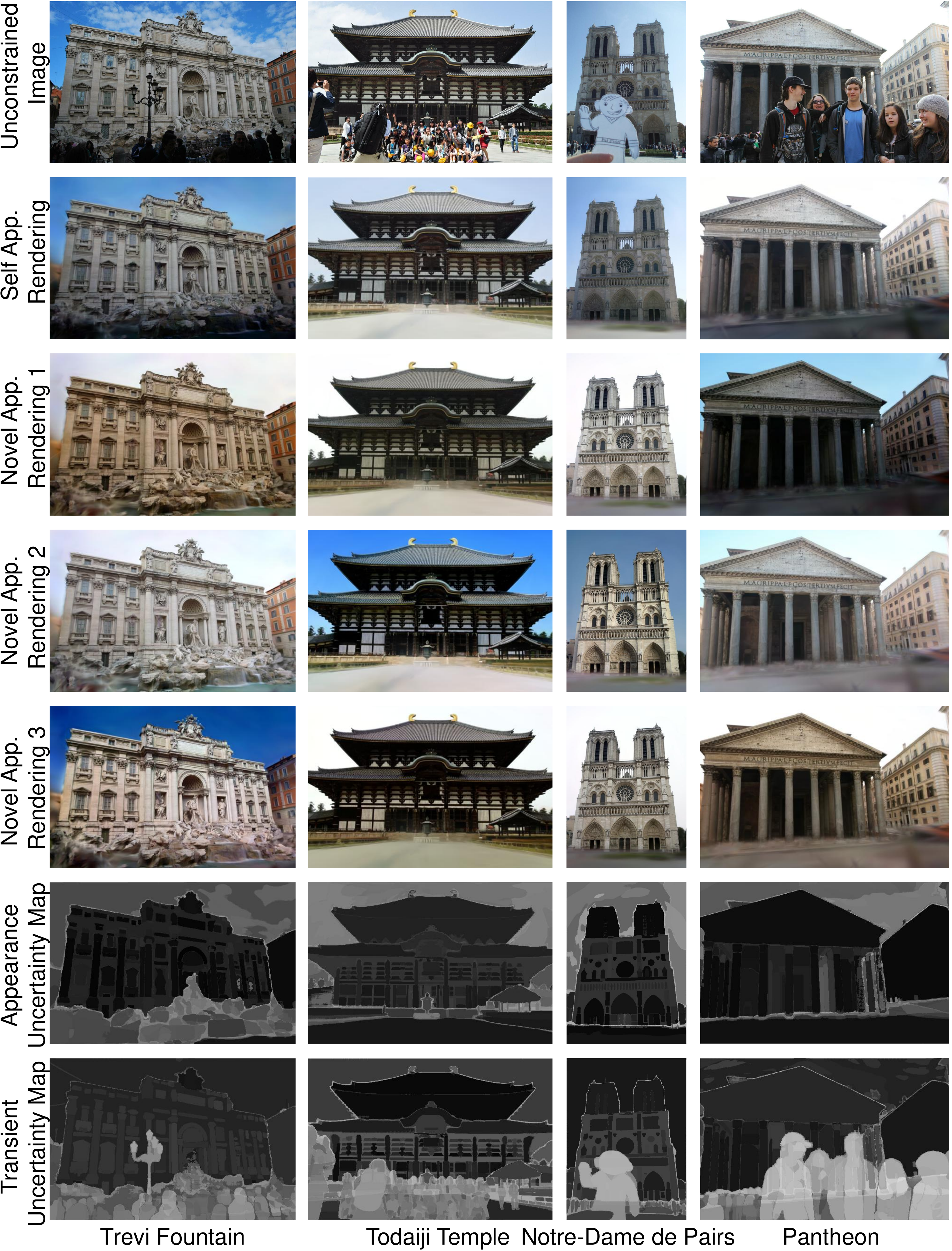}
    \caption{Visualization of self and novel appearance renderings, along with the appearance and transient uncertainty maps across different scenes. In the uncertainty maps, brighter colors represent higher uncertainty for each pixel.}
    \label{fig:transvis}
    \end{figure}

\begin{table}[htbp]
\caption{Ablation studies on the choice of novel appearance unconstrained images selection. Metrics are computed 5 times with different random seeds, reporting the average and fluctuation over 7 scenes in the proposed PT-OVS dataset.}
\resizebox{1.0\columnwidth}{!}{
\begin{tabular}{clccc}
\hline
  \multicolumn{2}{c}{}       & mIoU$\uparrow$           & mPA$\uparrow$   & mP$\uparrow$               \\ \hline
\multirow{4}{*}{\makecell{Novel App. \\ Selection}}&(1) w/ same App.      & 0.485±0.03 & 0.877±0.02          & 0.574±0.02            \\
                                   &(2) w/o  $\epsilon_d \& \epsilon_q$ & 0.561±0.13       & 0.947±0.07          & 0.652±0.11                \\
                                   &(3) w/o  $\epsilon_d$   & 0.620±0.09        & 0.958±0.06          & 0.745±0.13                \\
                                   &(4) w/o  $\epsilon_q$    & 0.563±0.17       & 0.947±0.12         & 0.652±0.17                 \\\hline
\multirow{4}{*}{\makecell{Number of \\Novel App.}}&(5) $N$=2      & 0.607±0.08      & 0.951±0.08       & 0.731±0.04                    \\
&(6) $N$=3      & 0.611±0.05      & 0.953±0.05       & 0.732±0.04                    \\
                            &(7) $N$=5          & \textbf{0.628}±0.04         & \textbf{0.961}±0.03       & 0.748±0.03\\
                            &(8) $N$=6      & \textbf{0.628}±0.04      & \textbf{0.960}±0.03       & \textbf{0.749}±0.02                    \\\hline
{Ours}&(9)  $N$=4    & {0.626±0.05}      & {0.960±0.03}         & {\textbf{0.749}±0.04}                   \\ \hline
\end{tabular}
}
\label{tab:ablation_study_app}
\end{table}
    Row 1 of \cref{ablation_study} shows that introducing multiple appearances enhancement improves segmentation accuracy.
    \cref{fig:transvis} visualizes the multiple appearance renderings across different scenes.
    Additionally, we conducted an ablation study to investigate whether the choice of novel appearances affects segmentation accuracy. As shown in \cref{tab:ablation_study_app}, we first perform a baseline experiment (row 1), where the three novel appearance unconstrained images are all set to the original unconstrained image. The segmentation accuracy in this setting is similar to that of the method without multi-appearance enhancement (row 1 of \cref{ablation_study}).
    In rows 2-4 of \cref{tab:ablation_study_app}, we relaxed the constraints on the selection of novel appearance unconstrained images. It can be observed that the rendering quality constraint $\selConsQuality$ for novel appearance rendering selection is particularly important. This is because in-the-wild scene radiance field reconstruction methods may fail to render photo-realistic images under certain appearances. When such images are used to extract CLIP features, they exhibit a significant domain gap compared to the features extracted from the original image. Additionally, the appearance embedding distance constraint $\selConsDistance$ is also crucial, as it prevents the appearance from becoming too similar, which would otherwise reduce performance back to the baseline (row 1).
    Row 2 of \cref{ablation_study} shows that performing multi-appearance fusion before training the autoencoder significantly reduces segmentation accuracy.
    Finally, we investigated the optimal number of novel appearances (Row 5-8 of \cref{tab:ablation_study_app}). We found that setting $N$ to 4 is the most reasonable choice. Although setting $N$ to 5 or 6 yields slightly better segmentation accuracy, it introduces additional storage overhead.
    
    \subsubsection{The Influence of Using Feature Uncertainty Map}
    Rows 3 and 6-8 of \cref{ablation_study} quantitatively demonstrate the important role of the feature uncertainty map in both the preprocessing and language field learning stages. Transient uncertainty map (TUM) effectively prevents the autoencoder from learning unnecessary features, such as those of transient occluders. In the language field learning stage, the introduction of the appearance uncertainty map (AUM) plays a more significant role in improving segmentation accuracy, as shown in rows 3 and 6-8 of \cref{ablation_study}. As shown in Rows 4–5 of \cref{ablation_study}, a lower $\tau_u$ results in degraded performance. While a larger $\tau_u$ yields comparable performance, it increases the training time of the uncertainty-aware autoencoder. Rows 6-7 of \cref{fig:transvis} visualize the appearance uncertainty map and the transient uncertainty map across different scenes. It can be observed that the transient uncertainty maps accurately represent occluders in unconstrained images, preventing the corresponding language features from being optimized during the autoencoder and language field learning processes. Meanwhile, the appearance uncertainty map identifies regions where language feature vary significantly across different appearances, enabling smoother optimization during training.
    
    \subsubsection{Analyze of the Post-Ensemble Strategy}
    We designed several ablations to ensemble the multiple appearance language score maps. Rows 9–13 of \cref{ablation_study} demonstrate the impact of different ensemble functions on segmentation accuracy. "ImgLvlMax. Ens." (row 8) indicates selecting the score map with the highest maximum value from the score maps as the final ensemble result. "PixMax. Ens." (row 9) and "PixAvg. Ens." (row 10) refer to ensemble functions that take the maximum and average values, respectively, at the pixel level from different score maps. "PixWeightedAvg. Ens." (row 11) is the method most similar to ours. It also employ a weighted average fusion approach, but it operates at the pixel level. We find that this approach can introduce more high-frequency noise into the final results, which decrease segmentation accuracy.
    
    \subsubsection{The Impact of the Autoencoder's Training Set}
    \begin{table}[htb]
\caption{Ablation studies on the training autoencoder with novel appearance language features. "ATT" is "autoencoder training time". Metrics are averaged over 7 scenes in the proposed PT-OVS dataset.}
\label{tab:ablation_autoencoder}
\resizebox{1.0\columnwidth}{!}{
\begin{tabular}{lcccc}
\cmidrule[\heavyrulewidth]{2-5}
            &ATT (h) $\downarrow$       & mIoU$\uparrow$                          & mPA$\uparrow$                           & mP$\uparrow$  \\\hline
w/\quad Novel App. Features           & 4.62    & 0.612   & 0.953  &0.733 \\
w/\quad Avg. Novel App. Features      & 9.83    & 0.502   & 0.890  &0.613 \\
w/o\: Novel App. Features  (Ours)     & $\mathbf{1.17}$    & $\mathbf{0.621}$   & $\mathbf{0.957}$  &  $\mathbf{0.745}$   \\ \hline
\end{tabular}}
\end{table}

    
    As shown in \cref{tab:ablation_autoencoder}, incorporating CLIP features from rendered novel appearance images into the autoencoder’s training set not only significantly increases training time (by approximately 3.5 hours) but also reduces segmentation accuracy (by about 1\% mIoU). We attribute this decline to the expanded domain of definition of scene-specific CLIP features, as the training set size roughly quadruples.
    The experimental setting in row 2 of \cref{tab:ablation_autoencoder} mirrors that in row 2 of \cref{ablation_study}. Integrating multi-appearance features prior to autoencoder training notably increases computation, as the feature granularity shifts from superpixel- to pixel-level. To alleviate this, we apply similarity-based clustering to the fused pixel-level features and average within clusters, reducing the training load in this experiment.
    
    \subsubsection{The Influence of the Quality of the Reconstructed Radiance Field}
    Does the quality of reconstructed radiance field affect scene understanding accuracy? To investigate this, we designed several baseline experiments:
    (a) WE-GS + Ours: This is the method uses in our method. We reconstruct the scene using WE-GS \cite{we-gs}, render novel appearances, and perform scene understanding based on the radiance fields reconstructed by WE-GS.
    (b) 3DGS$^\dagger$ + Ours: To investigate the impact of 3D geometric priors on scene understanding accuracy, we use vanilla 3DGS \cite{3dgs} for scene reconstruction and WE-GS for novel appearance rendering. This baseline allows us to assess how the initialization of the spatial point of 3D Gaussians affects the scene understanding accuracy.
    (c) GS-W + Ours: This method uses GS-W \cite{gs-w} for scene reconstruction and novel appearance rendering, followed by scene understanding based on the radiance field reconstructed by GS-W. This baseline helps us to evaluate the impact of different in-the-wild reconstruction methods on scene understanding accuracy.
    
\begin{figure}[!htb]
    \centering
    \includegraphics[width=1.0\columnwidth]{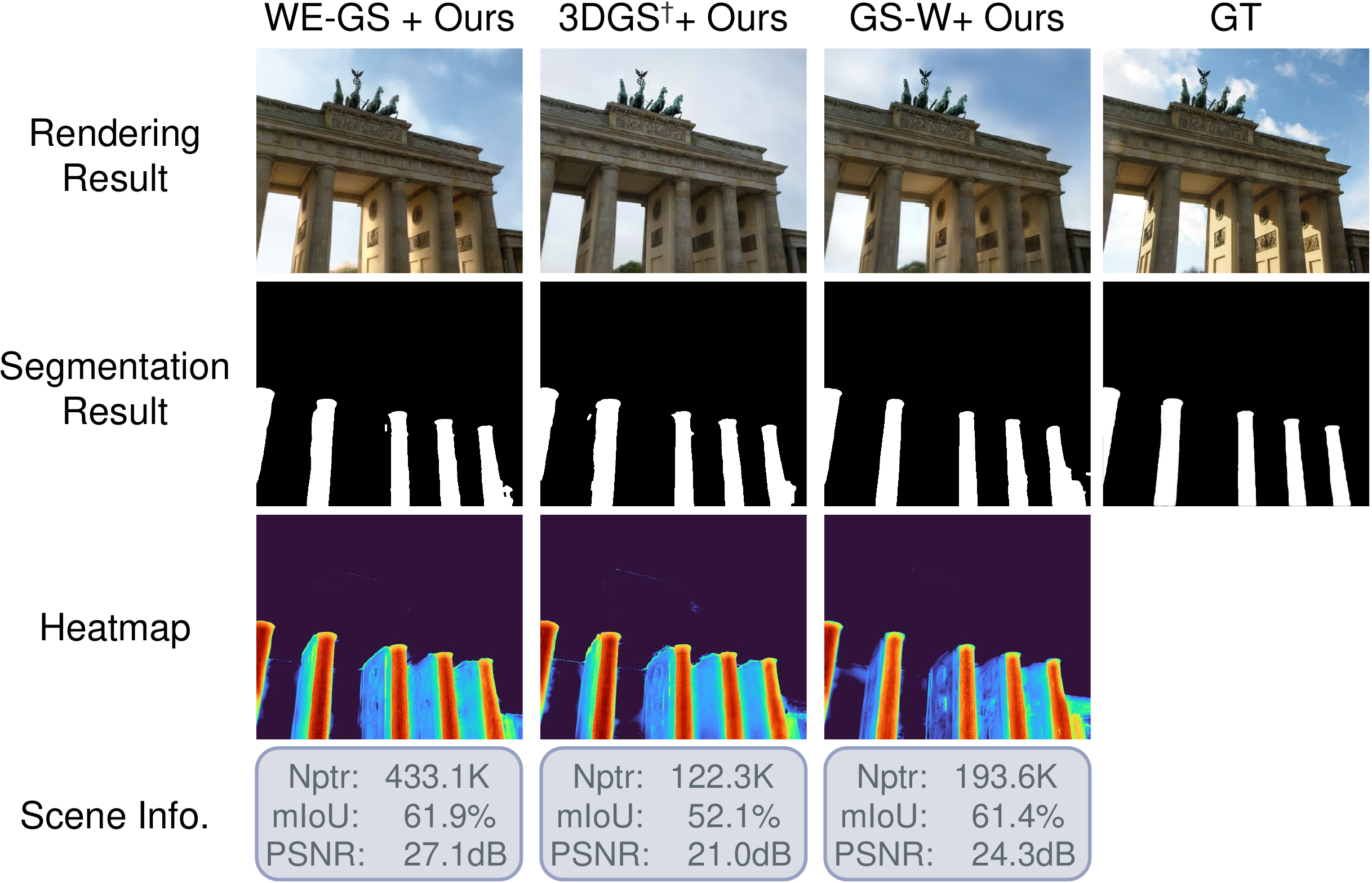}
    \caption{The ablation studies on the quality of reconstructed radiance field. WE-GS + Ours refers to our method. 3DGS$^\dagger$ + Ours uses WE-GS \cite{we-gs} for novel appearance rendering while initializing the language field with vanilla 3DGS. GS-W + Ours employs GS-W \cite{gs-w} for both novel appearance rendering and language field initialization. In the Scene Info., Nptr denotes the number of initialized 3D Gaussians in Brandenburg Gate scene, while mIoU and PSNR indicate the average mIoU on the PT-OVS test set and the average PSNR on the PT \cite{snavely2006photo} test set, respectively—both measured on the same scene.}
    \label{fig:fig_ablation_rfrecon}
    \end{figure}

    As shown in \cref{fig:fig_ablation_rfrecon}, vanilla 3DGS, which is used for static scene reconstruction with well-captured photo collections, produces a noisy point cloud with a large number of points (433.1K). This negatively impacts segmentation accuracy, resulting in a 9.8\% decrease in the mIoU metric.
    On the other hand, although the GS-W + Ours method yields slightly lower-quality radiance field in-the-wild reconstruction compared to WE-GS + Ours (with a PSNR difference of 2.8dB on the test set), the scene understanding performance remains comparable. This demonstrates the robustness of our method.

    \subsection{Application}
    We present three applications: interactive roaming with open-vocabulary queries, architectural style pattern recognition, and 3D segmentation and scene editing.
    \subsubsection{Interactive Roaming with Open Vocabulary Queries}
    The proposed method can seamlessly integrate with in-the-wild radiance field reconstruction approaches (e.g., WE-GS \cite{we-gs}, GS-W \cite{gs-w}). We develop an interactive system for users to remotely explore landmarks like the Trevi Fountain, with free-viewpoint roaming, lighting variations, and open-vocabulary queries. 
    We encourage readers to review the video results provided in the supplementary materials for additional details.
    
    \subsubsection{Architectural Style Pattern Recognition}
    \begin{figure}[!htb]
        \centering
        \includegraphics[width=1.0\columnwidth]{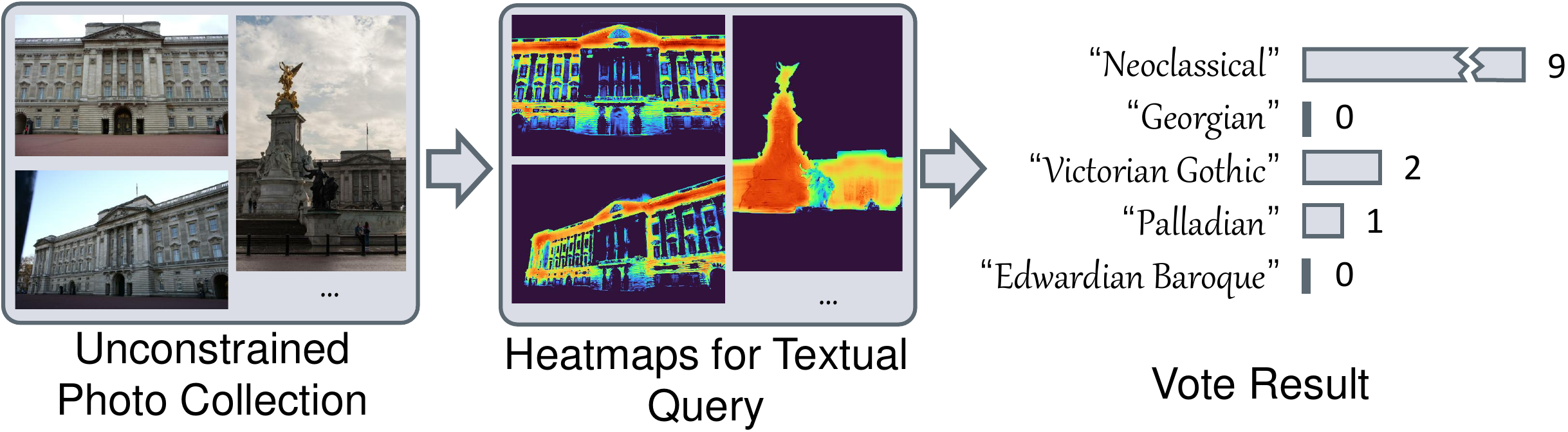}
        \caption{Application of architectural style pattern recognition.
        }
        \label{fig:fig_application_pr}
        \end{figure}
    As shown in \cref{fig:fig_application_pr}, our method is also effective for architectural style pattern recognition. Given a reconstructed language embedded 3DGS with our method, we query architectural style keywords (e.g., "Neoclassical", "Georgian", "Palladian") across multiple unconstrained images, generating multiple fused score maps. We take a "winner takes all" approach, selecting the highest value from different style score maps to vote for the corresponding style. Then, the predicted architectural style is determined by the most frequent vote. Compared to vanilla CLIP, our 3D language field-based method achieves more accurate recognition. For additional results, please refer to the supplementary materials.
    
    Additionally, we make an interesting discovery: the intermediate results of architectural style pattern recognition, specifically the heatmap of the fused score map, can be viewed as a visualization of "what makes Buckingham Palace a neoclassical building."

    \subsubsection{3D Segmentation and Scene Editing}
    
    Our method can further be applied to 3D segmentation and scene editing.
    
    After each 3D Gaussian has encoded the compressed language features, we decode them back into vanilla CLIP features at the 3D level. Using the same fusion approach as described in the method section, we perform open-vocabulary segmentation at the 3D Gaussian level instead of the 2D pixel level.
    \begin{figure}[!htb]
        \centering
        \includegraphics[width=0.95\columnwidth]{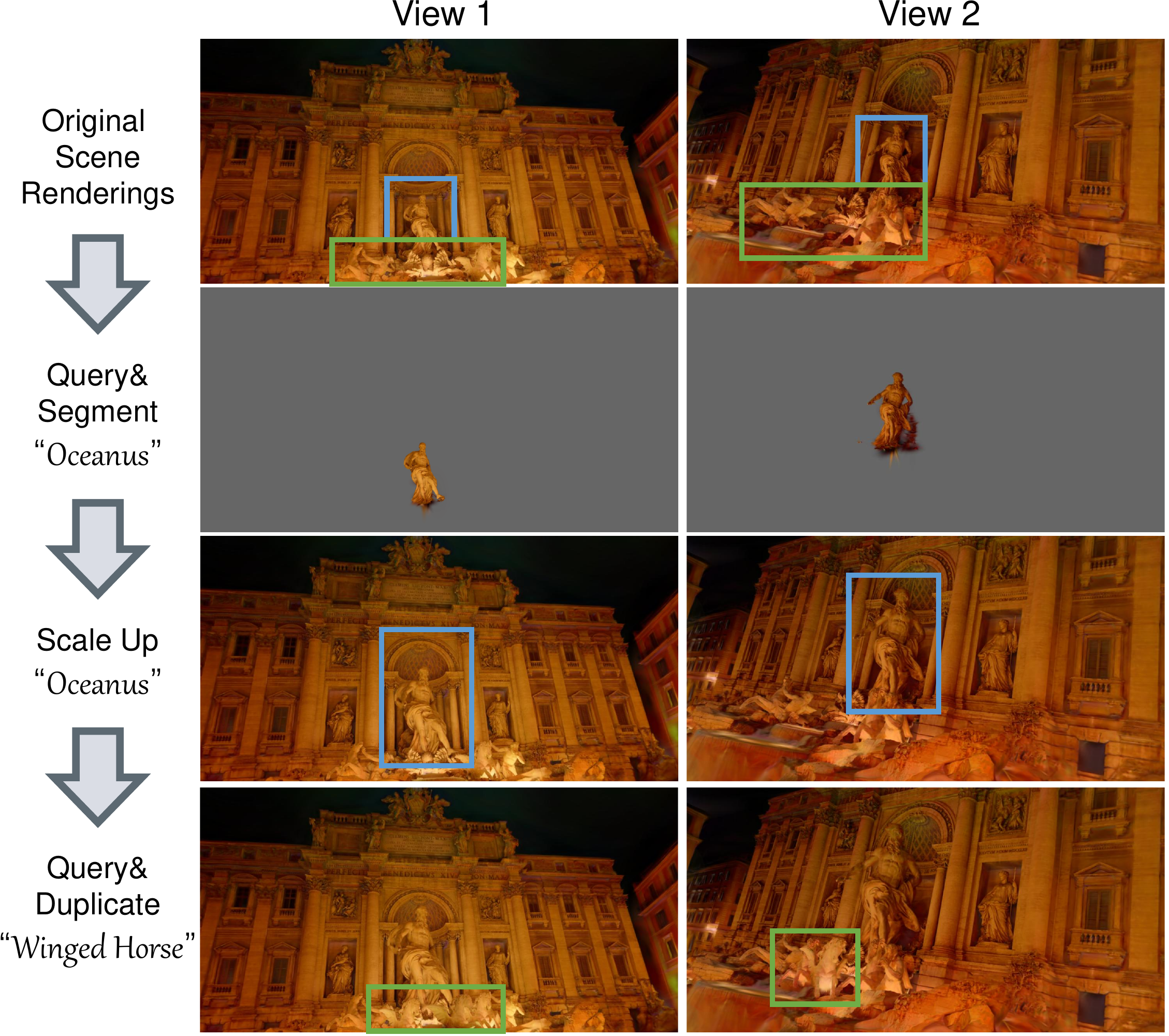}
        \caption{Application of 3D segmentation and scene editing. 
        }
        \label{fig:fig_application_editing}
        \end{figure}
    3D open-vocabulary segmentation has numerous applications in AR/VR, and we demonstrate the application of 3D scene editing.

    As shown in \cref{fig:fig_application_editing}, we query and edit "Oceanus" and "Winged Horse" in the scene of Trevi Fountain. Since our method can integrate seamlessly into the 3DGS editing and rendering pipeline, we use the open-source SuperSplat project \cite{SuperSplat} to scale, duplicate, and manipulate the selected 3D Gaussians, creating novel 3D scenes. Please refer to the supplemental materials for more details.

\section{Limitations}
\begin{figure}[!htb]
    \centering
    \includegraphics[width=0.9\columnwidth]{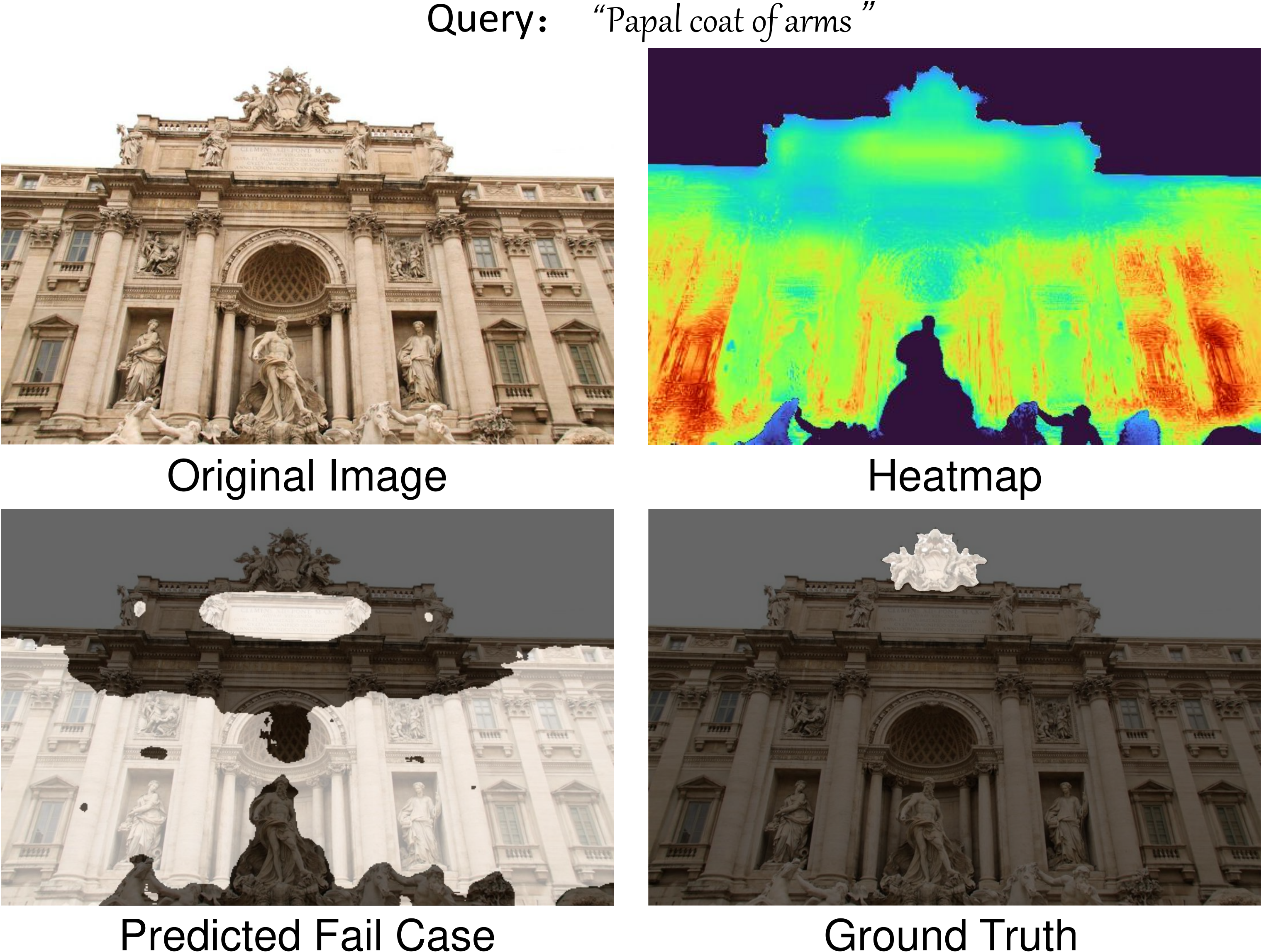}
    \caption{Failure Case. CLIP performs poorly on certain long-tail vocabulary, leading to failure in our method.}
    \label{fig:failcase}
    \end{figure}

While the proposed method outperforms previous approaches, it still has limitations.
Firstly, our method builds on CLIP features to link text and images. However, as illustrated in \cref{fig:failcase}, CLIP fails to handle certain long-tail vocabulary effectively. This issue could be mitigated by fine-tuning CLIP on more diverse, long-tail-oriented datasets or replacing it with stronger vision-language models such as BLIP \cite{li2022blip} or SigLIP \cite{zhai2023siglip}.
Secondly, although the multiple appearance enhancement strategy effectively handles 3D open-vocabulary understanding from unconstrained photo collections, it introduces additional parameters that need to be learned.

\section{Conclusion}
\label{sec:5_conclusion}
In this work, we take language embedded 3D Gaussian splatting into the wild.
We present a framework for reconstructing language fields from unconstrained photo collections using 3DGS.
Our framework incorporates several key components.
First, we propose a multi-appearance CLIP feature enhancement strategy for denoising and augmenting unconstrained language features.
Using the reconstructed radiance field, we render and extract multiple appearance language features from the same viewpoint and extract both the language feature transient uncertainty and appearance uncertainty map from them.
Next, we introduce a transient uncertainty-aware autoencoder, a multi-appearance language field 3DGS representation, and a post-ensemble strategy to compress, learn, and fuse multiple appearance language features.
Additionally, we introduce PT-OVS, a new benchmark dataset to evaluate open-vocabulary segmentation accuracy based on text queries for building components and historical contexts.
This paper enables the creation of semantically rich 3D content for AR/VR/MR from unconstrained photo collections, providing new insights into interactive querying, annotation, and the generation and editing of 3D content in AR/VR/MR environments.

\section*{Supplemental Materials}
\label{sec:supplemental_materials}
\setcounter{figure}{0}
\setcounter{table}{0}
\setcounter{section}{0}
\renewcommand\thesection{\Alph{section}}
\renewcommand\thetable{\Alph{table}}
\renewcommand\thefigure{\Alph{figure}}

In this supplementary material, we provide an in-depth explanation of our method, detailing additional the PT-OVS dataset in Sec. A, detailing implementation details in Sec. B, and demonstrating more experiment details and results in Sec. C.
\section{More Details of the PT-OVS dataset}
\label{sup_a}
\begin{figure}[!htb]
\centering
\includegraphics[width=1.0\columnwidth]{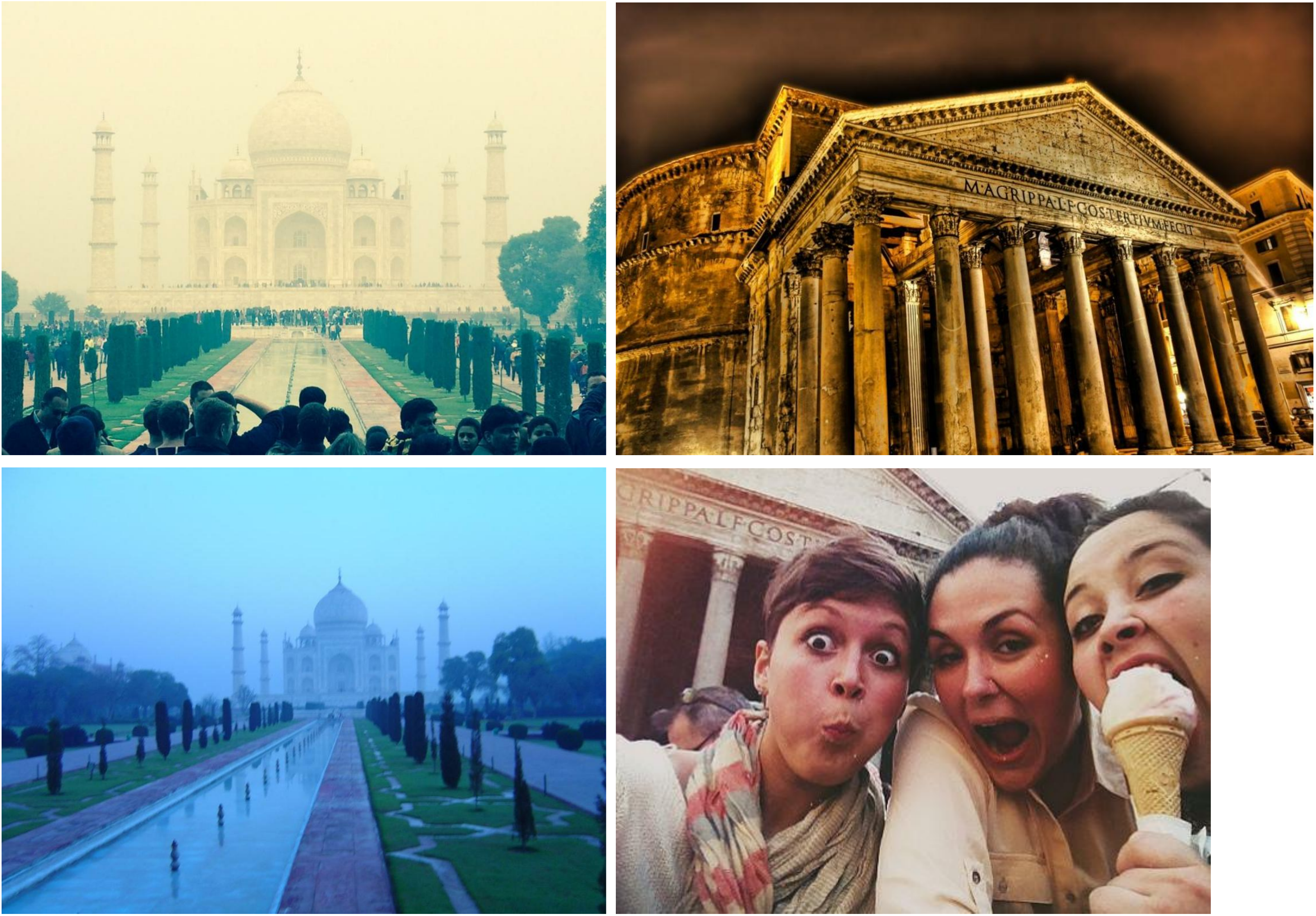}
\caption{Some filtered unconstrained images, which are typically of low quality or severely occluded.}
\label{fig:fig_sup_filter}
\end{figure}
\begin{table}[htb]
\caption{Number of training images, evaluation images, and query texts per PT-OVS scene.}
\label{tab:sup:tab_scene_info}
\resizebox{1.0\columnwidth}{!}{
\begin{tabular}{lccl}
\cmidrule[\heavyrulewidth]{2-4}
            &Train Img.      &Eval. Img. &textual queries   \\ \hline
Brandenburg Gate   & 763    & 16   & Column, Iron Cross, Bronze Sculpture    \\
Trevi Fountain                       & 1689    & 27  & Column, Oceanus, Winged Horses, Window, Reliefs      \\
Buckingham Palace                       & 1379   & 11   & Queen Victoria Memorial, Ionic Column, Pediment       \\
Notre-Dame de Paris                       & 2934    & 15   & Rose Window, Last Judgment       \\
Pantheon                      & 1308   & 11  &   Triangular Pediment, Obelisk of the Pantheon, Corinthian Column    \\
Taj Mahal                       & 1238    & 14   &  Onion Dome, Arch-shaped doorways, Jali Windows     \\
Todaiji Temple                      & 859    & 8   & Eave, Door, Column       \\\hline
\end{tabular}}
\end{table}

We select and download scenes from the train and validation set of the Image Matching Challenge Photo Tourism (IMC-PT) 2020 dataset \cite{imc-pt}. These scenes contain unconstrained images with provided relatively accurate camera intrinsic and extrinsic parameters, as well as initial point clouds. We aim to select landmarks from different locations, styles, and types. In total, we choose seven scenes: Buckingham Palace, Notre-Dame de Paris, Pantheon, Taj Mahal, Brandenburg Gate, Todaiji Temple, and Trevi Fountain. First, we need to filter the unconstrained photo collections by removing low-quality images, such as those with severe occlusions or excessively low resolution. NeRF-W \cite{nerf-w} has already provided filtered results for Brandenburg Gate and Trevi Fountain. For these two scenes, we follow the same selection as NeRF-W. For the other five scenes, we apply the following filtering criteria. First, we apply the Neural Image Assessment (NIMA) \cite{talebi2018nima} to perform an initial filtering of the unconstrained photo collections. Specifically, we remove unconstrained images with a NIMA score lower than 4. We further remove images in which transient objects occupy more than 80\% of the image area, as determined by a LSeg \cite{lseg} model trained on the ADE20K dataset. \cref{fig:fig_sup_filter} shows examples of unconstrained images that were filtered out, from the scenes of Pantheon and Taj Mahal.

\begin{figure}[!htb]
\centering
\includegraphics[width=1.0\columnwidth]{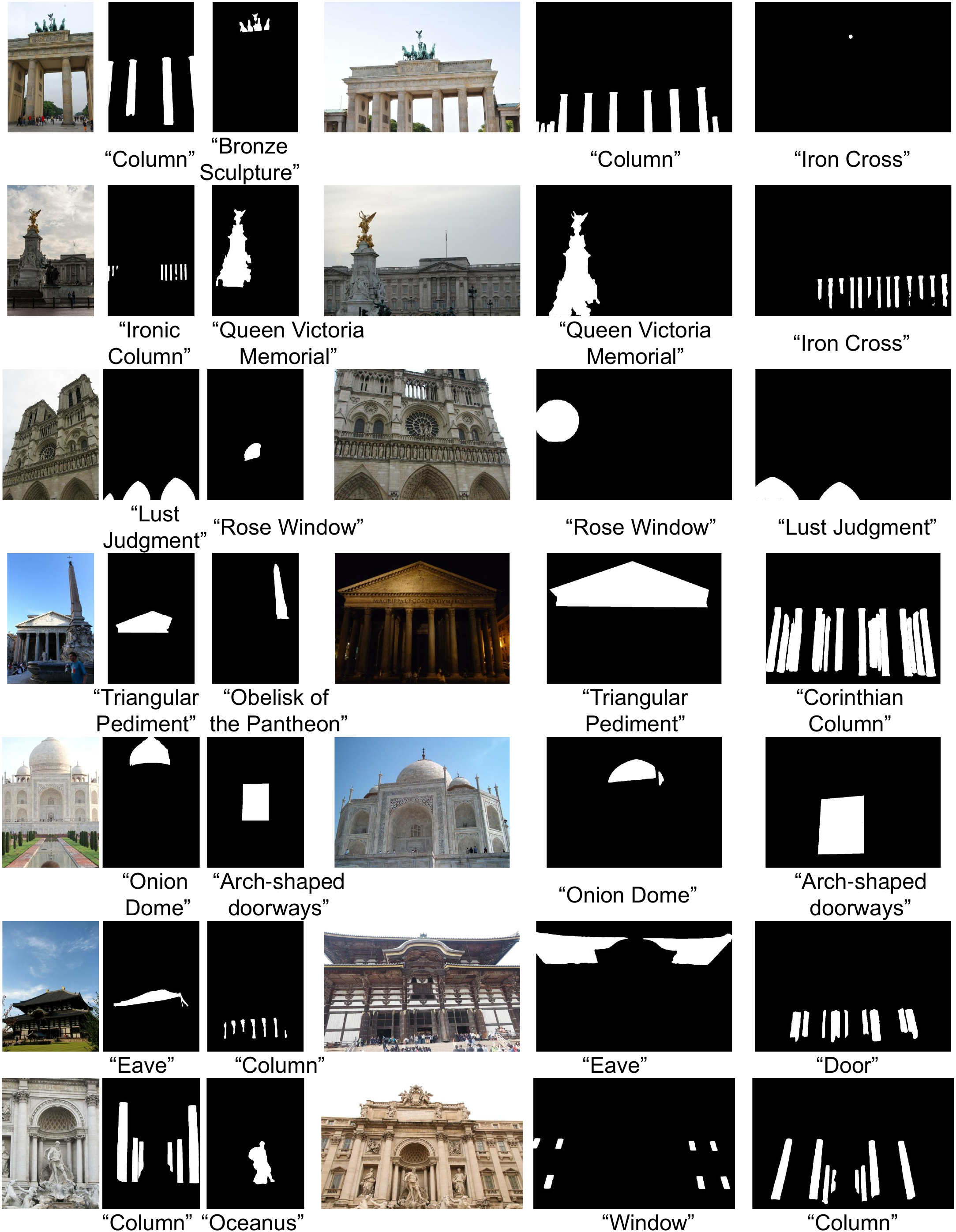}
\caption{Several evaluation images and their corresponding ground truth segmentation across seven scenes.}
\label{fig:fig_dataset}
\end{figure}
For evaluation, we carefully selected 10-20 high-quality unoccluded images from each scene. Annotations were made using the open-source image annotation tool ISAT \cite{ISAT_with_segment_anything}. For selecting open-vocabulary query texts, we mainly chose content that appeared in the scene’s description on Wikipedia. This includes commonly used architectural components (e.g., "Window"), architectural style terms (e.g., "Ionic Column"), proper nouns (e.g., "Oceanus"), and material nouns (e.g., "Bronze Sculpture"). \cref{tab:sup:tab_scene_info} provides a detailed breakdown of the number of unconstrained images used for training, the number of evaluation images, and the number of text queries for each scene.
It is worth noting that, to provide a comprehensive evaluation of open-vocabulary segmentation from unconstrained photo collections, we designed query texts that consider not only long-tail classes but also more general classes. For instance, we created long-tail queries such as "Ironic Column" as well as more common queries like "column". Additionally, we included distractor queries. For example, in the Pantheon scene, we used "Obelisk of the Pantheon" as the open-vocabulary query text instead of "Obelisk", to evaluate whether existing methods would segment the entire building instead of only extracting the obelisk.
\cref{fig:fig_dataset} presents several evaluation images along with their corresponding ground truth segmentation.

\section{More Implementation Details}
\subsection{Details of Multi-Appeaarance Pixel-Level Language Feature Extraction}
\subsubsection{Details of Radiance Field Reconstruction}
\begin{table}[htb]
\centering
\caption{Selected multi-appearance rendering IDs for all scenes.}
\label{tab:sup:tab_mulappid}
\resizebox{1.0\columnwidth}{!}{
\begin{tabular}{lrrr}  
\cmidrule[\heavyrulewidth]{2-4}
 & Novel App. 1 & Novel App. 2 & Novel App. 3 \\\hline
Brandenburg Gate   & 15080601\_1551250100.jpg    & 01738801\_5114523193.jpg   & 59826471\_8014732885.jpg   \\
Trevi Fountain     & 15457887\_10227170235.jpg   & 45182190\_511249303.jpg    & 80288369\_2336500045.jpg   \\
Buckingham Palace  & 04781012\_3416228976.jpg    & 11220321\_6429817645.jpg   & 42080522\_204096736.jpg    \\
Notre-Dame de Paris & 73272860\_3039447416.jpg    & 03158689\_7322662838.jpg   & 72005271\_4157221941.jpg   \\
Pantheon          & 00318896\_2265892479.jpg    & 02882184\_6968792622.jpg   & 04938646\_2803242734.jpg   \\
Taj Mahal        & 75255818\_297567547.jpg    & 76552970\_5828212829.jpg   & 86954812\_2533844894.jpg   \\
Todaiji Temple   & 08855480\_12135166146.jpg    & 10449189\_312833816.jpg    & 36907783\_8321442187.jpg   \\
\bottomrule
\end{tabular}}
\end{table}

For in-the-wild radiance field reconstruction, we re-implement a simplified version of WE-GS \cite{we-gs}. In WE-GS, a CBAM \cite{sa0_cbam} module is introduced to simultaneously predict the transient mask and appearance embeddings for each unconstrained image. The only difference in our implementation is that we do not use the channel and spatial attention blocks mentioned in WE-GS. Instead, we directly concatenate the predicted transient mask with the feature maps obtained through a CNN to achieve the attention mechanism, and then predict the appearance embeddings. All other network structures and hyperparameter settings follow the vanilla WE-GS across all 7 scenes.
\subsubsection{Details of Multi-Appearance Image Selection}
As shown in \cref{tab:sup:tab_mulappid}, we provide a detailed list of the multi-appearance rendering IDs selected for all scenes to ensure the reproducibility of the proposed method.

\subsubsection{Details of Uncertainty Map Cauculation}
For computing the appearance uncertainty map and transient uncertainty map, we apply a normalization step to address the issue of differing value ranges across different uncertainty maps. Specifically, for each scene, we record the maximum and minimum pixel values across all transient and appearance uncertainty maps. We then normalize each uncertainty map accordingly:

\begin{equation}
U_{i}^A = \frac{U_{i}^A - \min\limits_{i} U^A_i}{\max\limits_{i} U^A_i - \min\limits_{i} U^A_i}, \quad \forall I_i \in \mathbb{I},
\end{equation}
and
\begin{equation}
U_{i}^A = \frac{U_{i}^T - \min\limits_{i} U^T_i}{\max\limits_{i} U^T_i - \min\limits_{i} U^T_i}, \quad \forall I_i \in \mathbb{I}.
\end{equation}

\subsection{Details of Transient Uncertainty-Aware Autoencoder}
\begin{figure}[!htb]
\centering
\includegraphics[width=1.0\columnwidth]{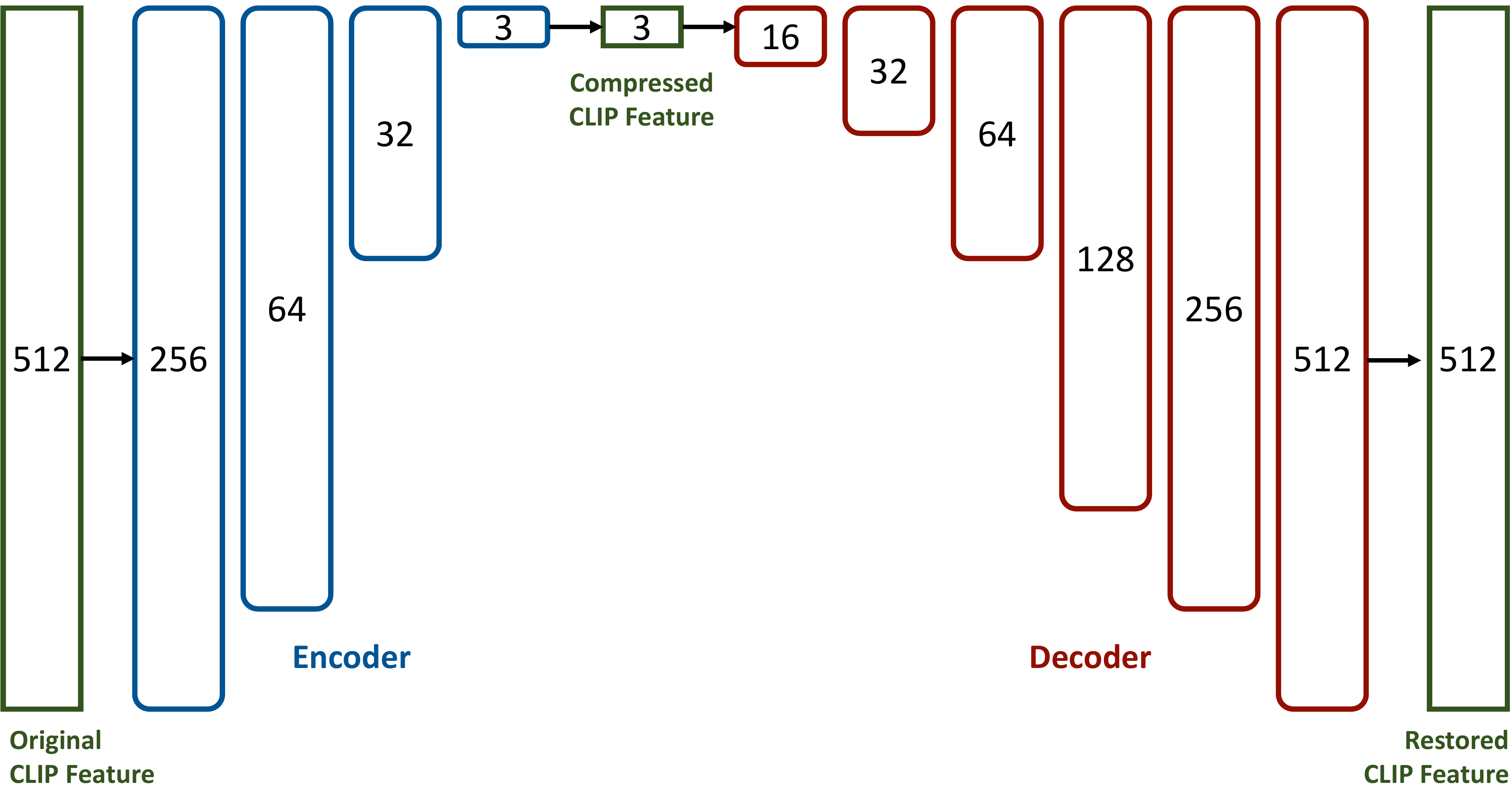}
\caption{The model architecture of autoencoder. Green rectangles represent input or output data, while blue and red rounded rectangles denote the encoder and decoder components, respectively. Each rounded rectangle represents an MLP, with numbers indicating the output dimensions.}
\label{fig:fig_aearch}
\end{figure}

Our autoencoder is implemented using MLPs, which compress the 512-dimensional CLIP features into 3-dimensional latent features. \cref{fig:fig_aearch} illustrates the network architecture of our model. We implement the model using PyTorch \cite{pytorch} and train it for 100 epochs with a learning rate of 0.0001 across all scenes.

\begin{figure}[!htb]
\centering
\includegraphics[width=1.0\columnwidth]{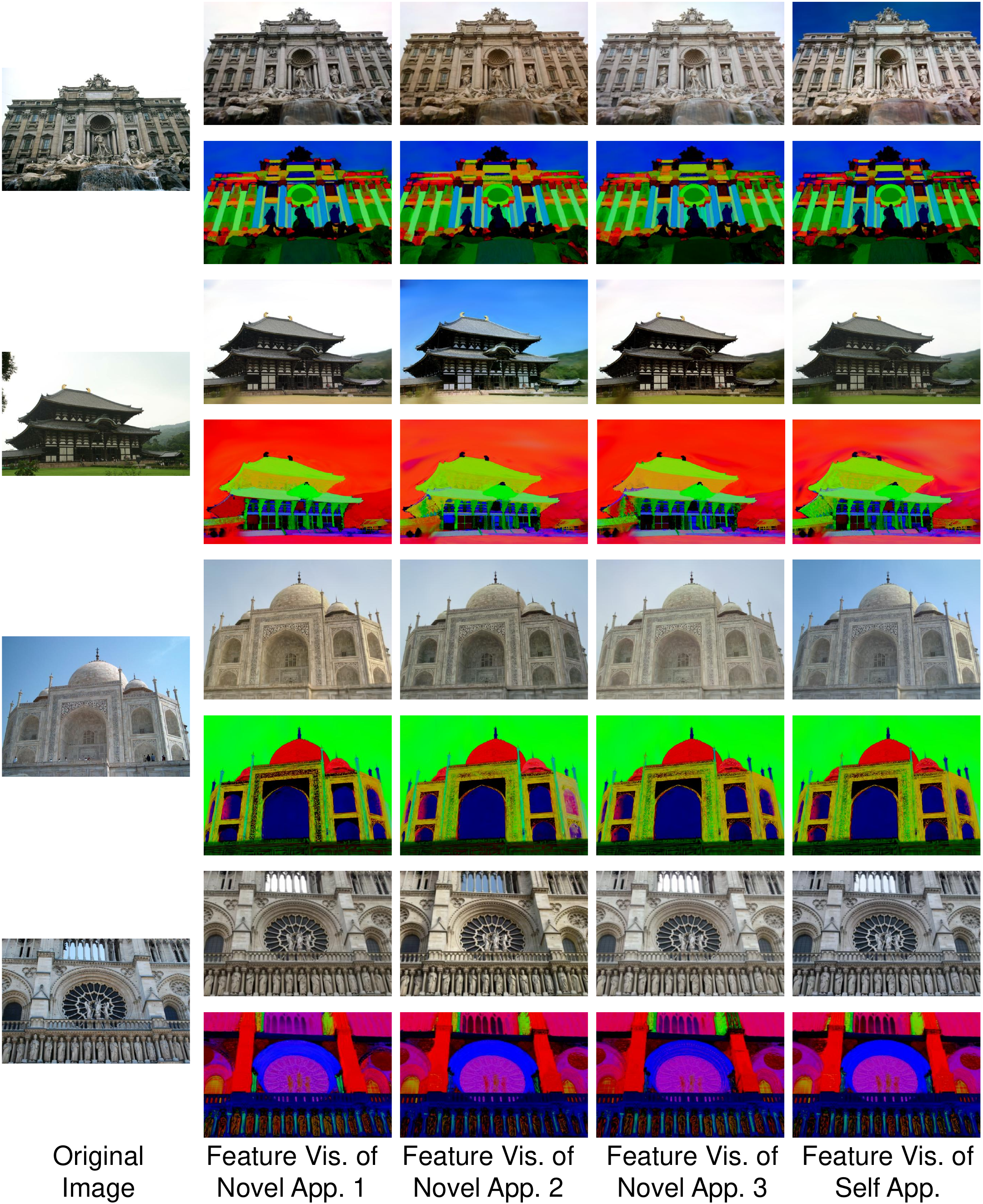}
\caption{Visualization of learned compressed language features from multi-appearance renderings. Since the scene-specific compressed language feature maps learned by MALE-GS are three-dimensional, we directly map these three dimensions to RGB for visualization.}
\label{fig:fig_sup_feat_vis}
\end{figure}
\begin{figure}[!htb]
\centering
\includegraphics[width=1.0\columnwidth]{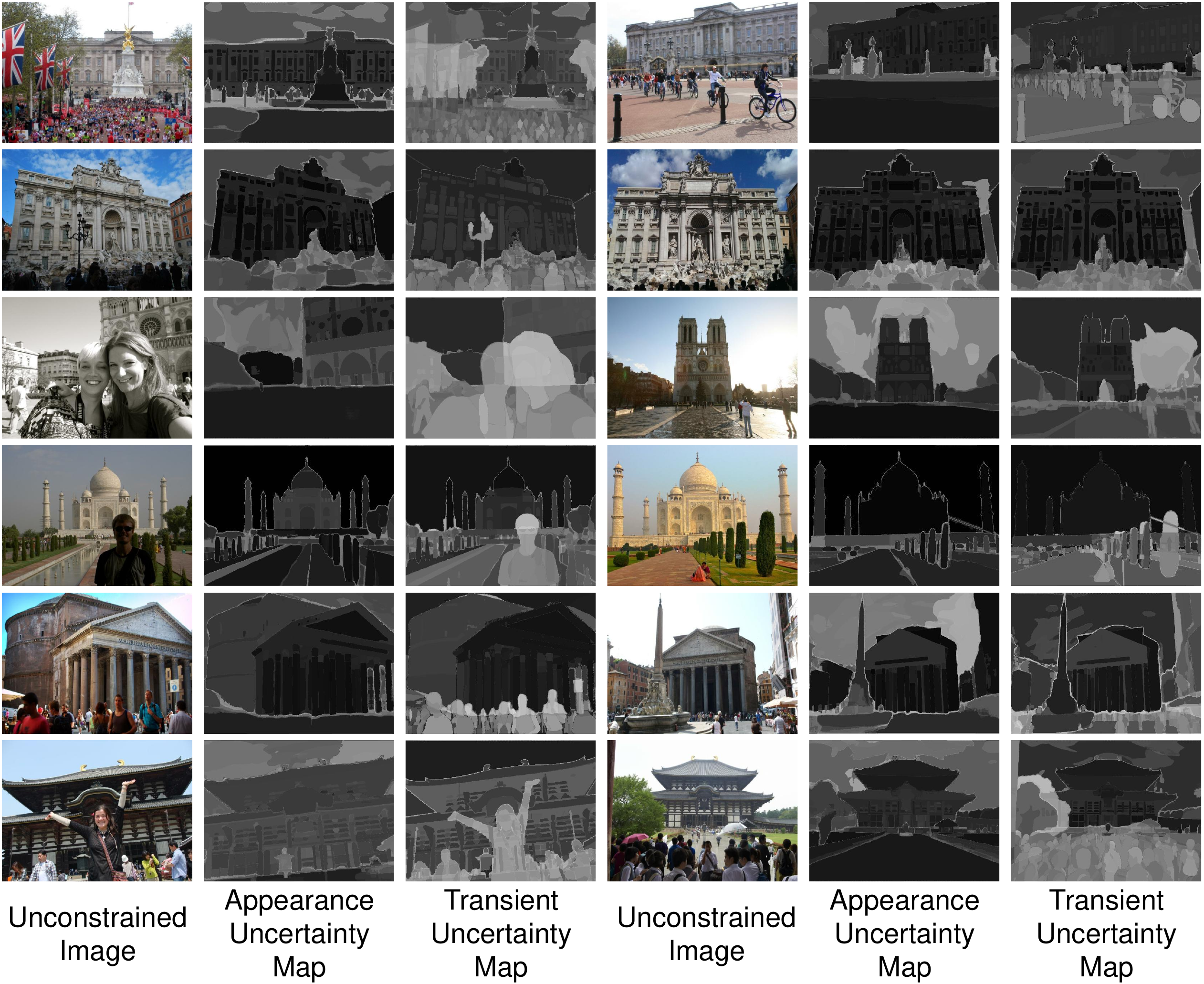}
\caption{More visualization results of the appearance uncertainty map and transient uncertainty map.}
\label{fig:fig_sup_moreuncertainlyvis}
\end{figure}
\begin{figure*}[!htb]
\centering
\includegraphics[width=2.0\columnwidth]{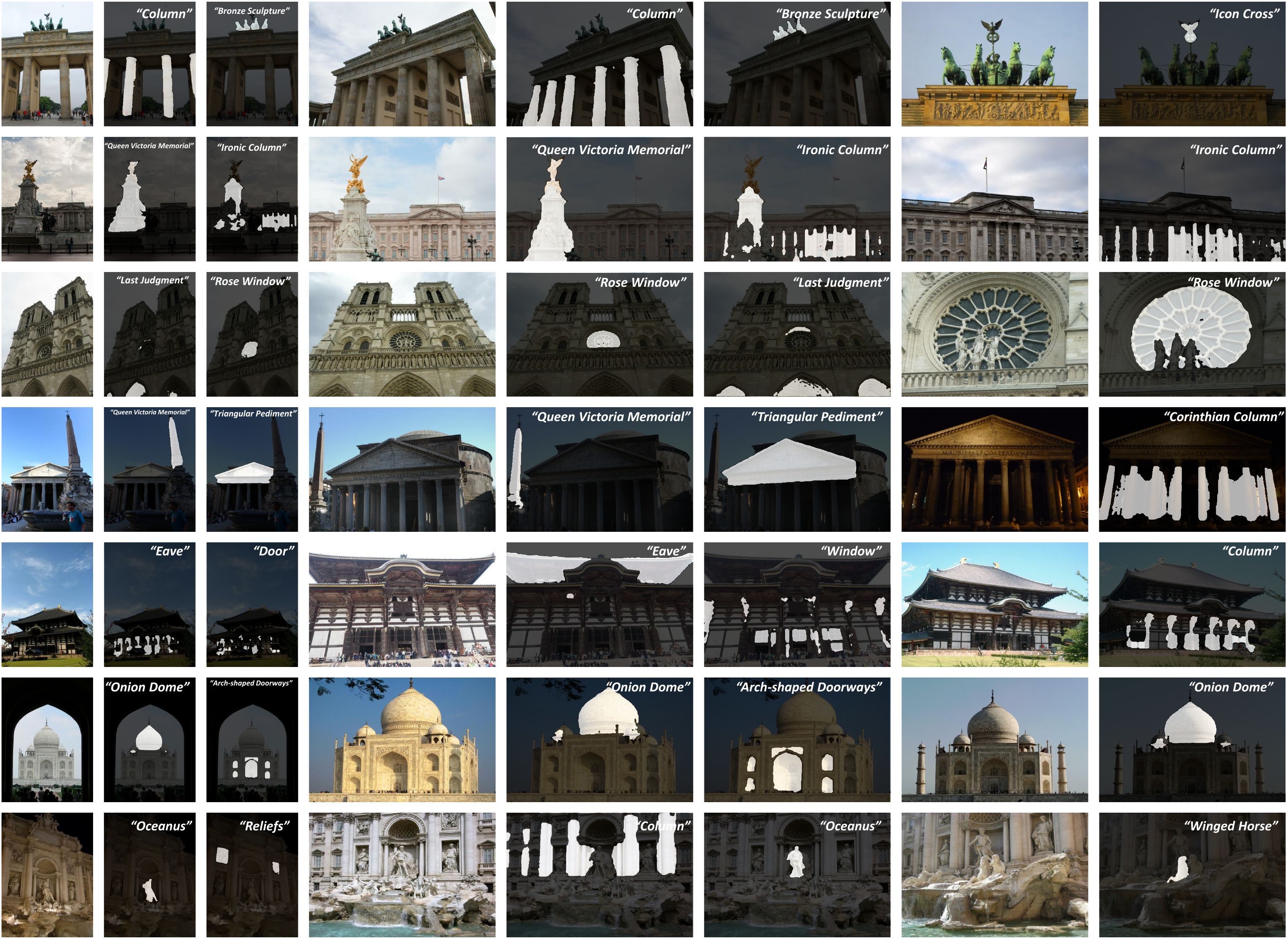}
\caption{More open-vocabulary segmentation results in the proposed PT-OVS dataset.}
\label{fig:fig_sup_moreresults}
\end{figure*}
\subsection{More Details of MALE-GS Representation and Optimization}
We implement the MALE-GS in Python using the PyTorch framework \cite{pytorch}, integrating custom CUDA acceleration kernels based on 3DGS's differentiable Gaussian rasterization \cite{3dgs}. We train the language features for 30,000 iterations with a learning rate of 0.0025.
As described in the main paper, we also use a hierarchical SAM-based method to generate pixel-level language features. We adopt the same strategy as LangSplat \cite{qin2024langsplat}, using SAM to define three semantic level: subpart, part, and whole. This results in three SAM segmentation maps and corresponding CLIP feature maps at these semantic level. To address the multi-scale nature of language features, we train three hierarchical MALE-GS. Given a query text, we evaluate the three MALE-GS models and select the one with the highest score, similarly with LangSplat.

\subsection{More Details of Post Ensemble and Open-vocabulary Querying}
The complete set of canonical texts consists of "object", "things", "scene", "sky", and "building".
To mitigate the influence of outliers, we apply a mean filter with a kernel size of 20 to smooth the relevancy maps, similar to LangSplat \cite{qin2024langsplat}.
\section{More Experiment Details and Results}
\begin{figure}[!htb]
\centering
\includegraphics[width=1.0\columnwidth]{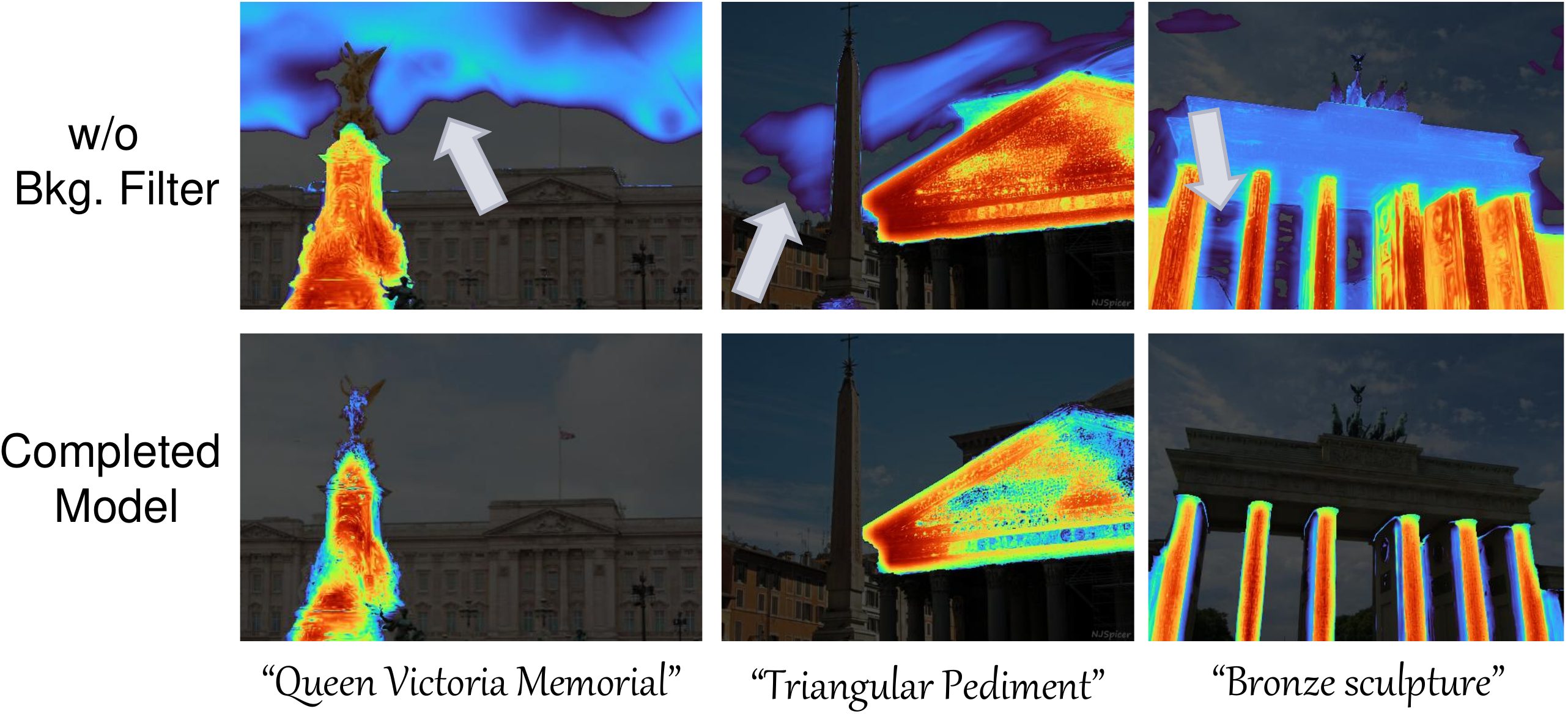}
\caption{Ablation studies on the background filter module. The heatmap depicts correlation: warmer colors represent stronger correlations, cooler colors indicate weaker correlations, and the absence of color signifies no correlation.}
\label{fig:ablation_sky}
\end{figure}
\subsubsection{The Influence of Using Background Filter}
As shown in row 9 of Tab. 3 of the main paper, the introduction of the background filter slightly improves segmentation accuracy. This is because we find that in outdoor scenes, existing methods using canonical phrases such as "object" and "stuff" as negative prompts cause the scoring mechanism to focus on the sky and other background areas. As illustrated in \cref{fig:ablation_sky}, the background filter significantly suppresses these regions, leading to more accurate segmentation results.
\subsection{Visualization of the Learned Multi-Appearance Language Features}

We demonstrate that CLIP features extracted from renderings with different appearances exhibit notable differences. As shown in \cref{fig:fig_sup_feat_vis}, the learned language feature maps after MALE-GS training vary significantly across appearances from the same viewpoint. Our proposed post-ensemble strategy effectively integrates these variations, leading to improved segmentation performance.

\subsection{More Qualitative Results}
\cref{fig:fig_sup_moreuncertainlyvis} shows additional visualizations of the language feature appearance and transient uncertainty maps.
 \cref{fig:fig_sup_moreresults} presents additional open-vocabulary segmentation results of our method. Due to the varying resolutions of the unconstrained images, some result images have been resized for clearer presentation.

\subsection{More Application Details and Results}
\subsubsection{More Details and Results of Interactive Roaming with Open-vocabulary Queries}
We implement the interactive roaming with open-vocabulary queries application by incorporating the proposed method into WE-GS. Since our open-vocabulary text query method is a downstream task of radiance field reconstruction, it can easily adapt to any in-the-wild radiance field reconstruction method, such as WildGaussians \cite{kulhanek2024wildgaussians}, Look at the Sky \cite{lookatthesky}, and GS-W \cite{gs-w}. In the specific implementation, we use in-the-wild radiance field reconstruction methods for appearance interpolation. Once the user inputs a query text, the system highlights the query results and allows interactive viewing from any viewpoint. We encourage readers to watch the supplementary video.

\begin{figure*}[!htb]
\centering
\includegraphics[width=2.0\columnwidth]{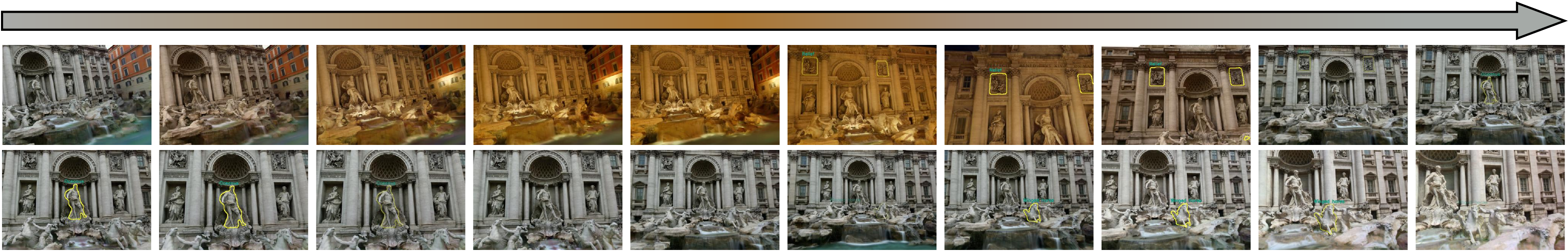}
\caption{Application of interactive roaming with open-vocabulary queries. The images, shown from left to right and top to bottom, are keyframes from a video generated by our method. Users can freely roam from any viewpoint, adjust lighting, and highlight architectural components with open-vocabulary queries. We encourage readers to watch our supplementary video.}
\label{fig:fig_application_roaming}
\end{figure*}

In the application case of the video supplemental material, we show the Trevi Fountain scene, involving appearance interpolation, text queries, and free-viewpoint roaming simultaneously. This system enhances the user's immersion in the process of understanding architectural components. It involves smooth interpolation between 5 different appearances and 3 different text queries, with the viewpoint zoomed in after the query results are retrieved.
For architectural component highlighting, we first obtain the segmentation results for the corresponding viewpoint using the proposed method. Then, we apply the Canny algorithm \cite{canny} to extract the edges of the segmentation mask. For text annotation, since we have the segmentation results, we can ensure that the annotation does not overlap with the queried architectural component, positioning it adjacent to the segmented region instead.
As shown in \cref{fig:fig_application_roaming}, when querying "Relief", the system accurately highlights and text-annotates the region, enabling visualization from any viewpoint and appearance. 
We encourage readers to review the video results provided in the supplementary materials for additional details.

\subsubsection{More Details and Results of Architectural Style Pattern Recognition}
\begin{figure}[!htb]
\centering
\includegraphics[width=1.0\columnwidth]{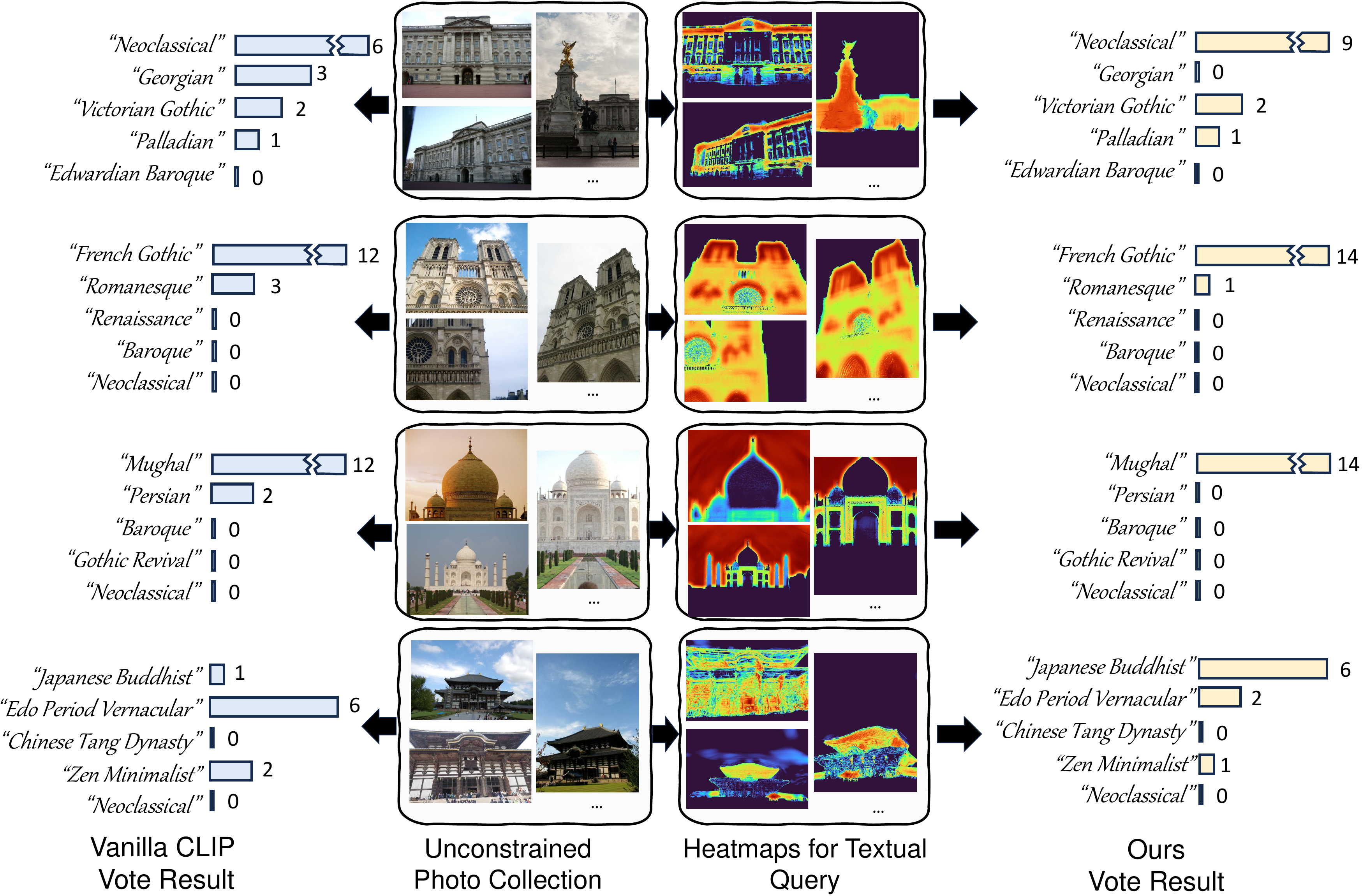}
\caption{More examples of architectural style pattern recognition applications.}
\label{fig:fig_sup_morepr}
\end{figure}
To enable architectural style pattern recognition with our method, similar to CLIP \cite{clip}, we predefine a set of candidate vocabularies and sequentially query them, selecting the most likely vocabulary as the pattern recognition result. we leverage large language model GPT-4 \cite{gpt4} to generate predefined candidate patterns. Given the scene name as input, GPT-4 outputs five candidate patterns. (Notably, we find that if the large language model directly predicts the architectural style of a scene, it may occasionally misidentify certain styles.) These five candidate patterns are then used in our method, where we randomly sample some unconstrained images and apply the "winner takes all" strategy mentioned in the main paper to vote for the most probable architectural style.

Fig. \ref{fig:fig_sup_morepr} presents additional results, where we also visualize heatmaps to illustrate "what makes a certain building a certain architectural style." Additionally, we report results obtained by directly using raw CLIP for pattern recognition. The results indicate that the proposed method achieves more accurate architectural style pattern recognition.

\subsubsection{More Details of 3D Segmentation and Scene Editing}
As described in the main paper, we decode the compressed language features at the per-Gaussian level and compare them with the text CLIP features to obtain a 3D mask, indicating which 3D Gaussians are selected. Once the 3D Gaussians are selected, we discard the multi-appearance language features within each Gaussian, converting the proposed MALE-GS representation into the vanilla 3DGS representation. This transformation allows seamless integration into any downstream applications.

We import the converted 3DGS into SuperSplat  and utilize its built-in transformation and scaling tools to manipulate the selected 3D Gaussians and merge them with the original scene. This enables the creation of an edited 3D scene that supports free-viewpoint roaming.

\subsection{Video Supplemental Material}
We provide a detailed introduction to our method and experimental results in the video supplemental material. We encourage the readers to watch the supplementary video for a clearer understanding of our approach.

\bibliographystyle{abbrv-doi}

\bibliography{2_reference}
\end{document}